\documentclass[preprint2]{aastex}
\usepackage{amssymb,amsmath} 

\newcommand{\Lya}{Ly$\alpha$}
\newcommand{\boldLya}{Ly$\boldmath\alpha$} 
\newcommand{\swift}{\textit{Swift}}
\newcommand{\betaOX}{\beta_\mathrm{OX}}

\shorttitle{Lyman-$\alpha$ emission from GRB host galaxies}
\shortauthors{Milvang-Jensen et al.}

\usepackage{natbib}
\begin{document}

\title{The Optically Unbiased GRB Host (TOUGH) survey. IV. Lyman-$\alpha$ emitters\altaffilmark{*}}
\altaffiltext{*}{Based on observations collected at
the European Southern Observatory, Chile,
as part of program 177.A-0591.}

\author{Bo Milvang-Jensen\altaffilmark{1},
Johan P. U. Fynbo\altaffilmark{1},
Daniele Malesani\altaffilmark{1},
Jens Hjorth\altaffilmark{1},
P{\'a}ll Jakobsson\altaffilmark{2},
and
Palle M{\o}ller\altaffilmark{3}
}

\altaffiltext{1}{Dark Cosmology Centre, Niels Bohr Institute,
University of Copenhagen,
Juliane Maries Vej 30, 2100 Copenhagen {\O}, Denmark}
\altaffiltext{2}{Centre for Astrophysics and Cosmology, Science Institute,
University of Iceland, Dunhagi 5, 107 Reykjav{\'i}k, Iceland}
\altaffiltext{3}{European Southern Observatory, Karl-Schwarzschild-Strasse 2,
85748 Garching bei M{\"u}nchen, Germany}

\email{milvang@dark-cosmology.dk}

\begin{abstract}
We report the results of a spectroscopic search for Lyman-$\alpha$ ({\Lya})
emission from gamma-ray burst (GRB) host galaxies.
Based on a well-defined parent sample (the TOUGH sample) of 69 X--ray selected
{\swift} GRBs, we have targeted the hosts of a subsample of 20 GRBs
known from afterglow spectroscopy to be in the redshift range $z$ = 1.8--4.5.
We have obtained spectroscopy using the FORS1 instrument at
the ESO Very Large Telescope
to search for the presence of {\Lya} emission from the host galaxies.
We detect {\Lya} emission from 7 out of the 20 hosts, with
the typical limiting $3\sigma$ line flux being
$8 \times 10^{-18}\,\mathrm{erg}\,\mathrm{cm}^{-2}\,\mathrm{s}^{-1}$,  
corresponding to a {\Lya} luminosity of
$6 \times 10^{41}\,\mathrm{erg}\,\mathrm{s}^{-1}$ at $z=3$.            
The {\Lya} luminosities for the 7 hosts in which we detect {\Lya} emission
are in the range (0.6--2.3)$\,\times 10^{42}\,\mathrm{erg}\,\mathrm{s}^{-1}$, 
corresponding to 
star-formation rates of
0.6--2.1$\,M_\sun\,\mathrm{yr}^{-1}$                                          
(not corrected for extinction).
The rest-frame {\Lya} equivalent widths (EWs) for the 7 hosts
are in the range 9--40\,{\AA}\@.
For 6 of the 13 hosts for which {\Lya} is not detected we place fairly strong
$3\sigma$ upper limits on the EW ($<20$\,{\AA}), while for others the EW 
is either unconstrained or has a less constraining upper limit.
We find that the distribution of {\Lya} EWs is inconsistent with
being drawn from the {\Lya} EW distribution of bright Lyman break
galaxies (LBGs) at the 98.3\% level, in the sense that the TOUGH hosts on average
have \emph{larger} EWs than bright LBGs.
We can exclude an early indication, based on a smaller, heterogeneous sample
of pre-{\swift} GRB hosts, that all GRB hosts are {\Lya} emitters.
We find that the TOUGH hosts on average have \emph{lower}
EWs than the pre-{\swift} GRB hosts, but the two samples
are only inconsistent at the 92\% level.
The velocity centroid of the {\Lya} line (where detected) is redshifted by
200--700$\,\mathrm{km}\,\mathrm{s}^{-1}$
with respect to the systemic velocity (taken to be the afterglow
redshift), similar to what is seen for LBGs, possibly
indicating star-formation driven outflows from the host galaxies.
There seems to be a trend between the {\Lya} EW and the
optical to X--ray spectral index of the afterglow ($\betaOX$), hinting
that dust plays a role in the observed strength and even presence of
{\Lya} emission.
\end{abstract}


\section{Introduction}

Due to their potential brightness at wavelengths ranging from radio to gamma-rays,
gamma-ray bursts (GRBs) and
their afterglows can be used as powerful astrophysical probes. They are
momentarily bright enough to be observed anywhere in the Universe, even if
located in the most dusty environments or at the highest redshifts
\citep[e.g.][]{Wijers_etal:1998}, but later fade away and allow a detailed
study of their environment. It is now established that long-duration
GRBs are associated 
with core collapse
supernovae \citep[e.g.,][]{Stanek_etal:2003,Hjorth_etal:2003,Woosley_Bloom:2006,Hjorth_Bloom:2012}
and hence star-formation.
GRBs thus can be used to
probe the star-formation density over most of the cosmic history from the
formation of the first stars to the present. The most spectacular example of
this is GRB\,090423 at $z = 8.2$
\citep{Tanvir_etal:2009,Salvaterra_etal:2009} representing a look-back time of
more than 95\% of the time since the Big Bang.

{\Lya}-emitting galaxies are in a state of active star-formation and most
likely contain little or no dust, since {\Lya} photons have a much higher
probability than other UV photons of being absorbed by dust due to resonant
scattering \citep{Adams:1972,Charlot_Fall:1993,Valls-Gabaud:1993}.
{\Lya} emission is hence sensitive to both the star-formation rate and the
dust content of GRB host galaxies. In addition, geometrical effects and the
kinematical state of the interstellar medium 
seem to be important
for the escape of {\Lya} photons
\citep[e.g.,][]{Giavalisco_etal:1996,Hayes_etal:2005,Verhamme_etal:2006,Laursen_etal:2009b}.

The first studies of {\Lya} emission from (pre-{\swift}) GRB host galaxies
indicated that {\Lya} emission seemed to be ubiquitous,
with 5 detections out of 5 possible
\citep{Fynbo_etal:2003a}\footnote{%
GRB\,971214 \citep{Kulkarni_etal:1998},
GRB\,000926 \citep{Fynbo_etal:2002},
GRB\,011211 \citep{Fynbo_etal:2003a},
GRB\,021004 \citep[][and references therein]{Moller_etal:2002}, and
GRB\,030323 \citep{Vreeswijk_etal:2004}.}.
This would be intriguing as only about 25\% of Lyman-break selected 
galaxies (LBGs) at similar redshifts have {\Lya} emission
with rest-frame equivalent width (EW) larger than 20\,{\AA}
\citep{Shapley_etal:2003};
this is also the case at higher redshifts
(\citealt{Douglas_etal:2010}, but see also \citealt{Stark_etal:2011}).
Another 6 GRB host {\Lya} emitters (all from {\swift} GRBs)
have been reported in the literature since then\footnote{%
GRB\,060714  \citep{Jakobsson_etal:2006},
GRB\,060926  \citep{Fynbo_etal:2009},
GRB\,061222A \citep{Perley_etal:2009},
GRB\,070110  \citep{Fynbo_etal:2009},
GRB\,071031  \citep{Fynbo_etal:2009}, and
GRB\,090205  \citep{DAvanzo_etal:2010}.}
(excluding the hosts reported in this work\footnote{This work reports
the detection of {\Lya} emission from 7 host galaxies, of which one
(GRB\,070110) was already identified as a {\Lya} emitter in the literature.}),
but there still has not been a systematic examination of the frequency of
{\Lya} emitters among GRB host galaxies. This is the aim of the present
work.

Given the effect of dust on {\Lya} photons,
possible explanations for an excess of {\Lya} emitters among GRB
host galaxies include \citep{Fynbo_etal:2003a}:
(i)~a preference for GRB progenitors to be metal poor as expected in
the collapsar model
(\citealt{Woosley_Heger:2006,Yoon_Langer:2005}; see also
 \citealt{Niino_etal:2009});
(ii)~an optical afterglow selection bias against dusty hosts;
(iii)~a higher fraction of {\Lya} emitters at the faint end of the high-redshift
luminosity function, where most GRB hosts are found;
(iv)~small-number statistics.
Using a well selected and more complete \swift\ sample we shall 
here address these issues.

The paper is structured in the following way.
The parent sample (TOUGH), the target selection,
the spectroscopic observations and the data reduction
are described in \S\ref{sec:targetsel_obs_datareduc}.
The {\Lya} detections and upper limits are presented in
\S\ref{sec:lya_fluxes_ews}.
The velocity offset of the {\Lya} emission with respect to 
the systemic velocity as given by the afterglow redshift
is discussed in \S\ref{sec:lya_velocity_offset}.
A comparison of the {\Lya} fluxes from afterglow and host spectra
is done in \S\ref{sec:lya_afterglow_host_comparison}.
In \S\ref{sec:summary_discussion} we discuss the results,
including how they relate to LBGs 
and to pre-{\swift} studies,
and how the observed {\Lya} emission is related to the
afterglow broad-band spectral index $\betaOX$,
and we summarize our findings.
Finally, the Appendix presents observations targeting the hosts of 3 GRBs
that are not part of the complete, well defined TOUGH sample discussed
in the main part of the paper.

We assume $H_0 = 70\,\mathrm{km}\,\mathrm{s}^{-1}\,\mathrm{Mpc}^{-1}$,
$\Omega_\mathrm{m}=0.3$, $\Omega_\Lambda=0.7$.
This only affects the {\Lya} luminosities and the derived
star-formation rates.
The reported magnitudes are on the Vega system,
with the exception of Fig.~\ref{fig:Lya_EW_comparison}. 

The reduced data from this work will be available from
ESO\footnote{%
http://archive.eso.org/
}
and from the TOUGH website\footnote{%
http://www.dark-cosmology.dk/TOUGH
}.

\section{Target selection, observations and data reduction}
\label{sec:targetsel_obs_datareduc}

\subsection{The TOUGH sample}
\label{sec:parent_sample}

This work is based on a parent sample named
The Optically Unbiased GRB Host (TOUGH) sample.
This sample of 69
{\swift} GRBs has several important features:
(i)~The selection criteria (see below) are designed to provide
an optically unbiased (X--ray selected) sample of long-duration GRBs;
(ii)~The selection criteria are also designed to increase the
prospects of prompt follow-up observations being successful;
(iii)~The sample has been the focus of an extensive prompt
follow-up campaign by our group (e.g.\ \citealt{Fynbo_etal:2009});
(iv)~The sample has been the focus of an extensive late-time
follow-up campaign targeting the host galaxies,
as 
reported in this series of papers
(\citealt{Hjorth_etal:2012,Malesani_etal:2012,Jakobsson_etal:2012};
this paper; \citealt{Kruhler_etal:2012:TOUGH5,Michalowski_etal:2012:TOUGH6}).

The sample selection criteria and their rationale are given in detail in
\citet{Hjorth_etal:2012}. 
They can be summarised as follows:
(1)~The burst should trigger the $\gamma$--ray imager BAT onboard {\swift};
(2)~Only long-duration bursts are considered;
(3)~An X--ray afterglow should be detected and
    the {\swift} XRT X--ray position should be
	made available within 12 hours from the trigger;
(4)~Milky Way extinction $A_V \le 0.5\,$mag;
(5)~Sun distance at the time of the GRB detection $> 55^\circ$;
(6)~No nearby bright stars (would complicate host galaxy observations);
(7)~Only bursts in the period 2005 March 1 to 2007 August 10 are considered;
(8)~Declination in the range $-70^\circ$ to $+27^\circ$
	(suitable for VLT observations);
(9)~The localization of the burst from the X--ray afterglow
	should be better than 2.0$''$ (90\% error radius)\footnote{%
	This includes using the revised UVOT-enhanced
	{\swift}-XRT positions \citep{Evans:2011,Evans_Osborne:2011}, which
        has had the effect of increasing the sample size from 68
        \citep[e.g.][]{Jakobsson_etal:2011:Como} to 69, with the additional
        burst being GRB\,060923B.}. 

Furthermore, observations targeting the host as part of the TOUGH
large program should be carried out at least 50 days after the GRB.

\subsection{Target selection}
\label{sec:targetsel}

The GRBs for the {\Lya} host galaxy spectroscopy studied here
were selected from the TOUGH sample
by applying the criterion that the (spectroscopic) redshift should 
be known and in the range 1.8 to 4.5. 
The lower limit comes from the atmospheric cut-off and the
sensitivity curve of the used CCDs, while the upper limit comes from
fringing in the CCDs used in some of the observing runs.

At the time of the target selection for the last run of the observing
campaign for the {\Lya} spectroscopy (\S\ref{sec:obs}),
the redshift status of the TOUGH sample
was as follows:
(a)~20 bursts met the $z$ = 1.8--4.5 criterion, and these were the 
ones observed, as listed in Table~\ref{tab:sample_obs}.
(b)~21 bursts had $z$ outside the range 1.8--4.5.
(c)~28 bursts did not have a secure, spectroscopic redshift determination.
Note that group~(b) included 5 redshifts obtained as part of the
TOUGH redshift campaign
\citep{Jakobsson_etal:2012} which were available before the
{\Lya} observing campaign ended.

For reference, the redshifts were subsequently revised for some bursts, and
redshifts became available for other bursts. 
As of February 2012, the split of the TOUGH sample into the three groups
based on redshift would be:
(a$'$)~27 hosts have $z$ = 1.8--4.5, of which 20 hosts are those
observed with FORS1 in the {\Lya} campaign and presented in this paper
(Table~\ref{tab:sample_obs}),
while 7 hosts do not have such FORS1 spectroscopy\footnote{%
These 7 additional bursts have redshifts from recent \mbox{X-shooter} host
spectroscopy \citep[][see also \S\ref{sec:betaOX}]{Kruhler_etal:2012:TOUGH5}. 
}.
(b$'$)~22 hosts have $z$ outside 1.8--4.5.
(c$'$)~20 hosts do not have a (secure, spectroscopic) redshift.

\begin{deluxetable}{lllrllccc}
\tabletypesize{\small}
\tablewidth{0pt}
\tablecaption{GRB sample and log of FORS1 {\Lya} host galaxy observations\label{tab:sample_obs}}
\tablehead{
\colhead{Name} &
\colhead{$z$} &
\colhead{Ref.} &
\colhead{$R_\mathrm{host}$} &
\colhead{Grism+filter} &
\colhead{CCD} &
\colhead{$T_\mathrm{exp}^\mathrm{total}$} &
\colhead{Seeing} &
\colhead{$A_V$} \\
\colhead{} &
\colhead{} &
\colhead{} &
\colhead{(mag)} &
\colhead{} &
\colhead{} &
\colhead{(hr)} &
\colhead{(arcsec)} &
\colhead{(mag)}
}
\startdata
 \object{GRB 050315}  & 1.9500                  & (1)      & $    24.4$                  & 600B         & new &   1.5                  & $ 0.84$ & 0.159 \\ 
 \object{GRB 050401}  & 2.8983                  & (2); (8) & $    26.1$                  & 600B         & old &   2.1                  & $ 0.76$ & 0.216 \\ 
 \object{GRB 050730}  & 3.96855                 & (3)      & $   >27.2$                  & 600V+GG435   & old &   1.8                  & $ 0.77$ & 0.168 \\ 
 \object{GRB 050820A} & 2.61469                 & (4)      & $    24.8$                  & 600B         & old &   2.6                  & $ 0.86$ & 0.147 \\ 
 \object{GRB 050908}  & 3.3467                  & (2)      & $    27.7$                  & 600B         & new &   2.2                  & $<1.1 $ & 0.083 \\ 
 \object{GRB 050922C} & 2.1992                  & (5)      & $   >26.3$                  & 600B         & old &   2.2                  & $<1.3 $ & 0.332 \\ 
 \object{GRB 060115}  & 3.5328                  & (2)      & $    27.1$                  & 600V+GG435   & old &   2.1                  & $ 1.23$ & 0.447 \\ 
 \object{GRB 060526}  & 3.2213                  & (2); (9) & $   >27.0$                  & 600B         & new &   2.2                  & $ 0.93$ & 0.221 \\ 
 \object{GRB 060604}  & 2.136                   & (6)      & $    25.5$                  & 600B         & new &   1.7                  & $<1.1 $ & 0.142 \\ 
 \object{GRB 060605}  & 3.773                   & (7)      & $   >26.5$                  & 600V+GG435   & new &   1.4                  & $<1.5 $ & 0.164 \\ 
 \object{GRB 060607A} & 3.0749                  & (2); (10)& $   >27.9$                  & 300V         & new &   2.2                  & $<1.4 $ & 0.096 \\ 
 \object{GRB 060707}  & 3.4240                  & (2)      & $    24.9$                  & 600V+GG435   & new &   1.4                  & $ 1.02$ & 0.071 \\ 
 \object{GRB 060714}  & 2.7108                  & (2)      & $    26.4$                  & 300V         & new &   1.5                  & $ 0.99$ & 0.261 \\ 
 \object{GRB 060908}  & 1.8836                  & (11)     & $    25.5$                  & 600B         & new &   1.5                  & $<0.9 $ & 0.099 \\ 
 \object{GRB 061110B} & 3.4344                  & (2)      & $    26.0$                  & 600B         & new &   2.2                  & $<1.0 $ & 0.127 \\ 
 \object{GRB 070110}  & 2.3521                  & (2)      & $    25.0$                  & 600B         & new &   1.5                  & $<1.3 $ & 0.048 \\ 
 \object{GRB 070506}  & 2.3090                  & (2)      & $    26.1$                  & 600B         & new &   1.5                  & $<1.2 $ & 0.130 \\ 
 \object{GRB 070611}  & 2.0394                  & (2)      & $   >27.0$                  & 600B         & new &   3.0                  & $ 0.88$ & 0.042 \\ 
 \object{GRB 070721B} & 3.6298                  & (2)      & $    27.5$                  & 300V         & new &   3.8                  & $<0.8 $ & 0.105 \\ 
 \object{GRB 070802}  & 2.4541                  & (2); (12)& $    25.1$                  & 600B         & new &   1.5                  & $ 0.78$ & 0.090 \\ 
\enddata
\tablecomments{
$R_\mathrm{host}$ is the $R$--band total magnitude (or $3\sigma$ upper limit)
of the host galaxy (before correcting for Galactic extinction)
from \citet{Malesani_etal:2012}.
CCD indicates which FORS1 CCD was used (cf.\ \S\ref{sec:obs}).
The seeing was measured using a Gaussian fit to stars that happened to be
in the slit in the combined spectrum. If no stars were available an upper
limit on the seeing was set as the size of the smallest galaxy in the slit.
$A_V$ is the Galactic extinction in the $V$--band from
\citet{Schlegel_etal:1998}, as obtained from NED\@.
The corresponding reddening is $E(B-V) = A_V/3.315$,
and the Galactic extinction in the $R$--band is $A_R = 2.673\,E(B-V)$.
}
\tablerefs{
(1)~\citet{Berger_etal:2005};      
(2)~\citet{Fynbo_etal:2009};
(3)~\citet{Chen_etal:2005};
(4)~\citet{Prochaska_etal:2007};
(5)~\citet{Piranomonte_etal:2008};
(6)~\citet{Fynbo_etal:2009}, but corrected for error (see text);
(7)~\citet{Ferrero_etal:2009};
(8)~\citet{Watson_etal:2006};
(9)~\citet{Thone_etal:2010};
(10)~\citet{Fox_etal:2008};
(11)~\citet{Fynbo_etal:2009}, but prompted by the {\Lya} redshift from this work (see text);
(12)~\citet{Eliasdottir_etal:2009}.
}
\end{deluxetable}

The redshifts of two of the hosts included in our {\Lya} campaign
warrant special mention.
GRB\,060604 was originally included because it had a tentative
afterglow redshift of $z=2.68$ proposed by \citet{Castro-Tirado_etal:2006}.
A subsequent re-reduction and analysis of the same data by
\citet{Fynbo_etal:2009} did not confirm that redshift.
Instead, an upper limit of $z \la 3$ was derived,
and a possible redshift of $z=2.124$ was suggested, based on a single
absorption line interpreted as \ion{Al}{2}.
We recently obtained an X-shooter host spectrum
\citep{Kruhler_etal:2012:TOUGH5} that
gives $z \approx 2.1359$ from H$\beta$, [\ion{O}{3}] and H$\alpha$.
This prompted the discovery of a wavelength calibration error in the
afterglow spectrum; the revised absorption-line redshift is $z \approx 2.1361$.
We will adopt the value $z = 2.136$.
GRB\,060908 was originally included because it had an
afterglow redshift of $z=2.43$ from \citet{Rol_etal:2006}.
Our spectroscopy (\S\ref{sec:obs})
gave a {\Lya} host emission redshift of $z=1.887$.
This prompted a re-analysis of the afterglow spectrum which did not find
evidence for $z=2.43$ but which did find an afterglow redshift of $z=1.8836$
reported by \citet{Fynbo_etal:2009},
matching our {\Lya} host redshift.

For the target selection for the TOUGH {\Lya} campaign
there was no requirement that the hosts should be detected in the
deep $R$--band imaging from the TOUGH imaging campaign \citep{Malesani_etal:2012}
or elsewhere. The statistics for the $R$--band imaging 
of the 20 observed systems with a secure redshift in the range 1.8--4.5
(Table~\ref{tab:sample_obs}) are:
14 hosts are detected in the $R$--band
(with $R$--band magnitudes in the range 24.4 to 27.7) and
6 hosts are not detected down to a typical $3\sigma$ limit of $R=27$.

All observed bursts have a detected optical afterglow.
This was not required, but is a consequence of the requirement of a known redshift
before the end of the observing campaign.
In all cases the redshift from the optical afterglow comes from
interstellar absorption lines,
both low-ionization metal lines (such as \ion{O}{1}, \ion{Si}{2}, and
\ion{C}{2})
and high-ionization lines (such as \ion{C}{4} and \ion{Si}{4}),
providing a good estimate of the systemic redshift of the host galaxy.
This is relevant for the interpretation of the velocity offset of
the {\Lya} emission line with respect to the afterglow redshift
(\S\ref{sec:lya_velocity_offset}).

\subsection{Observations}
\label{sec:obs}

Spectroscopic observations were completed using the FORS1 spectrograph
(cf.\ \citealt{Appenzeller_etal:1998}) 
on the Very Large Telescope (VLT) over the period May 2006 -- May 2008
in service mode.

The detector system of FORS1 was changed in April 2007, i.e.\
during the observing campaign (cf.\ Table~\ref{tab:sample_obs}).
The old system consisted of a single Tektronix CCD,
providing a pixel scale of 0.20$''\,\mathrm{px}^{-1}$.
The new system consists of two blue-optimized E2V CCDs,
providing a pixel scale of 0.25$''\,\mathrm{px}^{-1}$ when read out using
the default $2 \times 2$ binning as we did.
The two CCDs are mounted so that the small gap between them is in
the spatial direction; the gap has no practical consequences for our program.
Compared to the old detector system, the new detector system provides
a larger recorded wavelength range, a higher efficiency below 6000$\,${\AA},
and suffers from fringing above 6500$\,${\AA}\@.

All targets were observed using a 1.3$''$ wide longslit.
For most of the observing campaign, grisms 600B and 600V were
used depending on the redshift of the target (see Table~\ref{tab:sample_obs}).
Towards the end of the observing campaign the lower resolution but
higher throughput 300V grism was used instead of 600B for some targets.
The achieved wavelength range and spectral resolution for the different
grisms and detector systems are listed in Table~\ref{tab:wlrange_specres}.

\begin{deluxetable}{lcccccccc}
\tabletypesize{\small}
\tablewidth{0pt}
\tablecaption{Wavelength range and spectral resolution
\label{tab:wlrange_specres}}
\tablehead{
\colhead{Grism+filter} &
\colhead{Slit width} &
\multicolumn{2}{c}{\hrulefill\ Wavelength range \hrulefill} &
\colhead{FWHM} &
\colhead{$R$} &
\colhead{$c/R$} &
\multicolumn{2}{c}{\hrulefill\ Dispersion \hrulefill} \\
\colhead{} &
\colhead{} &
\colhead{old CCD} &
\colhead{new CCDs} &
\colhead{} &
\colhead{} &
\colhead{} &
\colhead{old CCD} &
\colhead{new CCDs} \\
\colhead{} &
\colhead{(arcsec)} &
\colhead{(\AA)} &
\colhead{(\AA)} &
\colhead{(\AA)} &
\colhead{} &
\colhead{($\mathrm{km}\,\mathrm{s}^{-1}$)} &
\colhead{(\AA\,px$^{-1}$)} &
\colhead{(\AA\,px$^{-1}$)}
}
\startdata
600B                           & 1.3 & 3300--5680 & 3030--6020                  & \phn6.5 & 690 & 430 & 1.17    & 1.47    \\ 
600V+GG435                     & 1.3 & 4510--6850 & 4250--7190                  & \phn6.2 & 920 & 330 & 1.15    & 1.44    \\ 
300V                           & 1.3 & \nodata    & 3000--8880\tablenotemark{a} &    14.1 & 420 & 710 & \nodata & 3.15    \\ 
600R+GG435\tablenotemark{b}    & 1.3 & 5090--7220 & \nodata                     & \phn6.5 & 950 & 320 & 1.05    & \nodata \\ 
\enddata
\tablenotetext{a}{When grism 300V as here is used without an order sorter filter,
a second order spectrum may be present for $\lambda > 6600\,${\AA};
this has no consequences for our program.}
\tablenotetext{b}{Grism 600R is not relevant for the main part of this paper,
only for the Appendix.}
\tablecomments{
The FWHM values were measured from the [\ion{O}{1}]$\lambda5578\,${\AA} skyline
in the combined science frames.
The resolving power $R = \lambda/\mathrm{FWHM}$ values were computed at
the central wavelength of the available wavelength range.
Note that the listed values of FWHM and $R$ only correspond to the spectral resolution
of the obtained host galaxy spectrum if the observed spatial profile of the galaxy
(i.e.\ the intrinsic profile convolved with the seeing) was flat over the slit. For a peaked spatial profile
the obtained spectrum will have a better spectral resolution (i.e.\ smaller FWHM and
larger $R$).
}
\end{deluxetable}

The targets, which were generally too faint to be seen in an acquisition
image, were put in the slit using one of two methods, both involving a nearby
reference star. Either the position angle of the slit was set so that
the slit would go through the reference star and the target, or
the reference star was put in the slit, after which an offset was
applied to the telescope to put the target in the slit.
The required position angle or offset was computed based on
the $R$--band detection of the host or, for the hosts that were undetected
or only marginally detected in our $R$--band host imaging,
based on the position of the afterglow.

Each target was observed for a total exposure time of 1.4--3.8\,hr
(see Table~\ref{tab:sample_obs}), split into 4--8 individual exposures.
The individual exposures were dithered along the slit.
GRB\,060714 was observed twice, since the first observation was
obtained in poor transparency conditions.
In the first observation the galaxy continuum was not detected,
whereas in the second observation it was.
The first observation was done using grism 600B, whereas the
second observation was done using the more efficient grism 300V
and with a slightly larger exposure time.
In the analysis we will only use the data from the second observation.

Additionally, three bursts not in the TOUGH sample 
were observed. These are discussed separately in the Appendix.

\subsection{Data reduction}
\label{sec:datareduc}

Data reduction was performed mainly using IRAF\footnote{%
IRAF is distributed by the National Optical Astronomy Observatories,
which are operated by the Association of Universities for Research in
Astronomy, Inc., under cooperative agreement with the National Science
Foundation.}.
The individual frames were bias subtracted and flat fielded.
Cosmic ray events were removed using LACosmic \citep{vanDokkum:2001}.
A 2D wavelength calibration was established for each grism and observing night
(based on arc frames obtained the following day)
and applied to the corresponding science frames.
Vacuum wavelengths were used.
The wavelength calibration was verified using the few strong skylines
available, in particular [\ion{O}{1}]$\lambda$5578.89. 
The shifts in pixels in the spatial direction between the individual
science frames were computed in two ways: simply using the requested shifts
stated in the so-called observation blocks (OBs), and
using other objects (preferably stars) on the slits.
The two derivations of shifts in all cases agreed to within 1~pixel ($\approx 0.2''$).

The individual wavelength calibrated science frames were shifted in the 
spatial direction and combined (averaged).
Usually no weights were used, but 
for GRB\,050820A and GRB\,060604,
the signal in other objects in the spectrum varied in a way indicating
a variable transparency, and here weights were used.
For GRB\,050820A the weights were set to the flux of the reference star
which was in the slit; the weights were in the range 0.25--1.00 when normalized
to the largest value.
For GRB\,060604 no reference star was centered in the slit. Based on other
objects in the slit, weights in the range 0.67--1.00 were assigned.

To represent the uncertainties for each pixel in the
2D science spectra, we calculated 2D sigma spectra 
based on photon noise and readout noise.
The error (in ADU) in the given pixel is given by
\begin{equation}
\sigma = \sqrt{
\frac{f}{g \cdot n} + \left(\frac{\mathrm{RON}}{\sqrt{n}}\right)^2
}
\label{eq:sigma_spectra}
\end{equation}
where
$f$ are the counts in ADU in the given pixel in the
2D science spectrum before sky subtraction, 
$g$ is the gain (conversion factor) in $e^-$/ADU for a single image,
$n$ is the number of single images that were averaged in the combination,
and
$\mathrm{RON}$ is the read-out noise (in ADU) for a single image.

The spectra and sigma spectra were flux calibrated based on sensitivity functions
derived from 30 standard star observations and reference data from
\citet{Hamuy_etal:1992,Hamuy_etal:1994} and \citet{Oke:1990}.
The spectra and sigma spectra were corrected for atmospheric extinction.
The extinction curve for La Silla was used
\citep{Tug:1977,Schwarz_Melnick:1993},
since no extinction curve was available for Paranal
at the time of the reduction. 
A comparison between the La Silla curve and
the Paranal FORS1 broad-band extinction coefficients \citep{Patat:2003}
shows a very good agreement \citep{Milvang-Jensen_etal:2008}.

The spectra and sigma spectra were finally corrected for Galactic extinction
\citep{Cardelli_etal:1989,ODonnell:1994} using $R_V = 3.1$,
with $E(B-V)$ taken from \citet{Schlegel_etal:1998}.
Table~\ref{tab:sample_obs} lists the used values. 

It should be noted that the requested observing conditions were
to have a transparency that was better than or equal to thin cirrus.
This means that the spectra were not necessarily obtained under
photometric conditions, and thus that some of the derived fluxes may be
affected by thin cirrus.
Any such effect is mitigated by the rescaling of the spectra based on
the photometry (\S\ref{sec:lya_fluxes_ews}).

\subsection{Subtraction of neighboring objects in the spectra}
\label{sec:Subtraction_of_neighboring_objects}

\begin{figure*}[htbp]
\makebox[\textwidth]{\includegraphics[width=1.00\textwidth,bb=34 625 547 751]{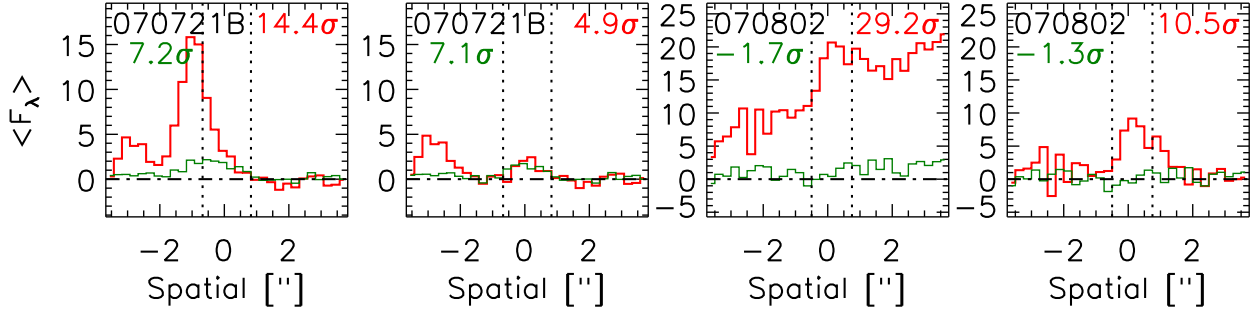}}
\caption[]{
Spatial profiles of the GRB\,070721B and GRB\,070802 spectra
before (1st panel) and after (2nd panel) a neighboring
object was fitted and subtracted in the 2D spectrum.
The red thick curve shows the continuum on the red side of {\Lya}, and
the green curve shows {\Lya}.
The subtraction of the neighboring object is evident from the red thick curve.
The rest of the features of this plot are explained in the caption of
Fig.~\ref{fig:spatialprofile_montage_sample}.
\label{fig:spatialprofile_montage_testNS}
}
\end{figure*}

The spectra of two of the hosts were substantially contaminated by
a neighboring object, which we fitted and subtracted in the 2D spectra.
For GRB\,070721B ($z=3.6$) a foreground galaxy ($z=3.1$) 1$''$ away
was fitted as a Gaussian in the spatial direction and a polynomial
in the wavelength direction.
For GRB\,070802 the wings of a bright star in the slit 18$''$ away
was fitted as a polynomial in both directions.
The fits were performed using MPFIT \citep{Markwardt:2009,More:1978}.
Figure~\ref{fig:spatialprofile_montage_testNS} shows the spatial profiles
before and after the subtraction of the neighboring objects.
The subtraction makes the derived continuum flux densities of the hosts
be 3.2 times (GRB\,070721B) and 2.6 times (GRB\,070802) smaller
and much more in line with what the photometry predicts (cf.\ below).
The EWs of {\Lya} (or upper limits thereof) become
larger by the same factors.
The {\Lya} fluxes, which are continuum-subtracted (cf.\ \S\ref{sec:lya_fluxes_ews}),
are practically unaffected.

\section{Results}
\label{sec:results}
\subsection{{\boldLya} detections and upper limits}
\label{sec:lya_fluxes_ews}

We first describe the measurement of continuum flux densities in the spectra.
We then compare these with the photometry and derive an average
correction for slit loss and extraction aperture loss.
We finally describe the measurement of {\Lya} fluxes and
EWs from the spectra.

To measure the continuum flux density in the spectra,
apertures on the blue and red
side of {\Lya} were defined as follows. In rest-frame wavelength
the apertures were 175$\,${\AA} wide and located such that a guard interval
of $\pm6\,${\AA} (corresponding to $\pm1500\,\mathrm{km}\,\mathrm{s}^{-1}$)
centered on {\Lya} placed at the afterglow redshift was excluded. 
In the spatial direction the continuum apertures coincided with the
{\Lya} apertures (see below). The width of 175$\,${\AA} was the maximum value
that was covered by all spectra, and this value also allowed
an accurate measurement of the continuum flux density.
Uncertainties on all measured fluxes and flux densities
were calculated by propagating the
individual uncertainties from the 2D sigma spectra (\S\ref{sec:datareduc}).

\begin{figure*}[htbp]
\makebox[\textwidth]{\includegraphics[width=1.00\textwidth,bb=8 570 564 745]{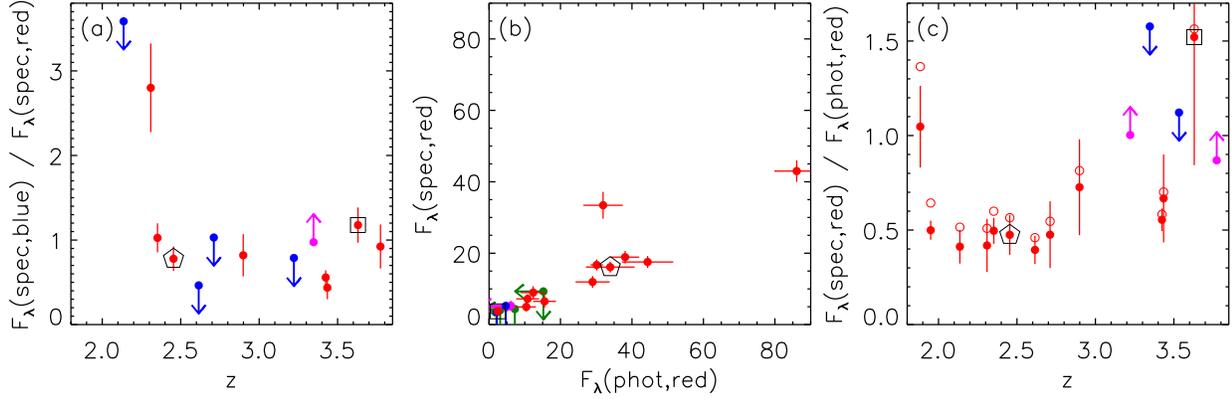}}
\caption[]{
Comparison of continuum flux densities.
Panel (a) compares the continuum flux densities measured in the spectra
on the blue and red side of {\Lya}.
Panels (b) and (c) compare the continuum flux densities measured in the
spectra with those 
from the photometry 
(filled points: $\beta = -1.5$, open points [panel~c only]: $\beta = -1.0$,
cf.\ Eq.~\ref{eq:F_lambda_phot_red}),
both corresponding to the red side of {\Lya}.
The units of the flux densities in panel~(b) are
$10^{-20}\,\mathrm{erg}\,\mathrm{cm}^{-2}\,\mathrm{s}^{-1}\,\mathrm{\AA}^{-1}$.
The blue and red windows are centered at 1122\,{\AA} and 1309\,{\AA}
and have widths of 175\,{\AA}, all rest-frame.
The two systems where a neighboring object was fitted and subtracted
are marked: GRB\,070721B (box) and GRB\,070802 (pentagon).
The discrepant point in panel~(a) at $z=2.3$ is due to the blue window having
very little signal and being dominated by a systematic error
from a time-variable pattern in the bias,
which has no effect on the derived {\Lya} properties presented in this paper
as they are based on the continuum measured in the red window in the spectra
(see text).
The symbol colors in the panels simply reflect the symbol type:
red = detections, blue = upper limits, magenta = lower limits,
green = upper limits in both the $x$ and $y$ direction.
The spectroscopic fluxes in this figure have not been multiplied by a factor
of 2.0 to correct for slit losses and aperture losses.
\label{fig:contfluxden_compare}
}
\end{figure*}

In Fig.~\ref{fig:contfluxden_compare}(a) we compare the spectroscopy-based
blue and red continuum flux densities.
Of the 20 hosts in the sample,
the red continuum is detected for 14 hosts, while
the blue continuum is only detected for 9 hosts,
all at $\ge3\sigma$ confidence;
note that two of the blue non-detections are outside the plotted range
in Fig.~\ref{fig:contfluxden_compare}(a).
One host (GRB\,050908, $z=3.3$, the magenta lower limit)
is just above the $3\sigma$ detection limit
the blue ($3.5\sigma$) but just below it in the red ($2.2\sigma$),
which is plausible given the redshift and the sensitivity curve of
the used grism (600B)\@.
The seemingly discrepant point at $z=2.3$ (GRB\,070506)
is due to a time-variable pattern in the bias
which we were unable to fully remove or quantify in terms of the error bars.
This only has a noteworthy effect in the far blue where the sensitivity is
very low and where the flux calibration therefore corresponds to a large
amplification.
This issue has no effect on the reported {\Lya} line properties,
as we have adopted the continuum flux densities measured in the red window
in the spectra as those used to subtract the (small) continuum contribution
from the measured flux in the {\Lya} aperture and to calculate the EWs.

In order to investigate the absolute flux scale of the
fluxes and flux densities extracted from the spectra, we use our
FORS2 $R$--band host imaging \citep{Malesani_etal:2012}.
The imaging was obtained largely under photometric conditions, and
the photometry was calibrated using \citet{Landolt:1992}.
The derived magnitudes are total magnitudes,
obtained either from using a large aperture, or from using a smaller
aperture combined with an aperture correction.
The correction was computed by analyzing the brighter host galaxies,
and adopting as uncertainty the observed scatter.

Starting from the total $R$--band host magnitude
$R_\mathrm{host}$, we correct for Galactic extinction ($A_R$) to
obtain a flux density at the observed-frame wavelength of the $R$--band of
\begin{equation}
F_{\lambda,R} = f_{\lambda,\mathrm{eff},R}^\mathrm{Vega} \times 10^{-0.4(R_\mathrm{host}-A_R)} \enspace ,
\end{equation}
where
\begin{equation}
f_{\lambda,\mathrm{eff},R}^\mathrm{Vega} = 2.15 \times 10^{-9}\,\mathrm{erg}\,\mathrm{cm}^{-2}\,\mathrm{s}^{-1}\,\mathrm{\AA}^{-1}
\label{eq:F_lambda_conversion_factor}
\end{equation}
is the conversion factor for Cousins $R$ from \citet{Fukugita_etal:1995};
practically the same factor would be obtained from
\citet{Blanton_Roweis:2007}\footnote{%
For Bessel $R$, \citet{Blanton_Roweis:2007} find
$m_\mathrm{AB}-m_\mathrm{Vega} = 0.21$\,mag and
$\lambda_\mathrm{eff} = 6442$\,{\AA},
which corresponds to a conversion factor
(Eq.~\ref{eq:F_lambda_conversion_factor})
of $2.16 \times 10^{-9}\,\mathrm{erg}\,\mathrm{cm}^{-2}\,\mathrm{s}^{-1}\,\mathrm{\AA}^{-1}$.
}.
We then extrapolate this flux density from the effective wavelength of
the $R$--band \citep[6410\,{\AA},][]{Fukugita_etal:1995}
to the observed-frame center wavelength of our red window,
$(1+z)\,1309\,\mathrm{\AA}$,
by assuming an $F_\lambda \propto \lambda^\beta$ spectrum, giving
\begin{equation}
F_\lambda(\mathrm{phot,red}) = F_{\lambda,R} \times \left( \frac{(1+z)\,1309\,\mathrm{\AA}}{6410\,\mathrm{\AA}} \right)^\beta \enspace ,
\label{eq:F_lambda_phot_red}
\end{equation}
where $\beta$ is the rest-frame UV spectral slope.
We will use $\beta = -1.5$ as a representative value
\citep[e.g.,][]{Shapley_etal:2003,Douglas_etal:2010,Finkelstein_etal:2011}.
Figure~\ref{fig:contfluxden_compare}(c) shows the effect of $\beta$
as function of redshift, since
in addition to the adopted value of $\beta = -1.5$ (filled points),
the case of $\beta = -1.0$ is illustrated (open points).

The spectroscopic flux densities show a good correlation with those
from the photometry. This is shown in
Fig.~\ref{fig:contfluxden_compare}(b), where the two are plotted
against each other, and in Fig.~\ref{fig:contfluxden_compare}(c) where
their ratio is plotted against redshift.
12~hosts have both a continuum detection (at $3\sigma$) in the spectra
on the red side of {\Lya}
and a detection in the $R$--band host imaging (at $2\sigma$).
These hosts are shown as red points in
Fig.~\ref{fig:contfluxden_compare}(c). 
2~hosts have a continuum detection in the spectra
but not in the imaging; these are
GRB\,060526 and  
GRB\,060605      
and are shown as magenta lower limits in Fig.~\ref{fig:contfluxden_compare}(c).
2~hosts conversely do not have a continuum detection in the spectra
but have a detection in the imaging; these are
GRB\,050908 and  
GRB\,060115      
and are shown as blue upper limits in Fig.~\ref{fig:contfluxden_compare}(c).

The ratio of spectroscopic to photometric flux density
shown in Fig.~\ref{fig:contfluxden_compare}(c)
has a median value 
of
0.50 for the photometry extrapolated using $\beta = -1.5$. 
We attribute the fact that this median ratio is lower than 1 to
slit losses (i.e.\ flux falling outside the slit) and
extraction aperture losses (i.e.\ flux falling outside the used extraction
apertures, as defined below),
as well as possibly a small amount of thin cirrus affecting the observations
(\S\ref{sec:datareduc}).
We use this result to derive an approximate global scaling factor
of $1/0.50 = 2.0$ that we apply to all the flux calibrated science
and sigma spectra (\S\ref{sec:datareduc}).
This has the effect of making the derived 
{\Lya} fluxes (or upper limits) and their uncertainties a factor of 2.0 larger,
while the EWs are unaffected by this procedure\footnote{%
In reality the {\Lya} EWs may in some cases be affected by slit loss,
namely if the {\Lya} and continuum emission differ in term of
spatial distribution. We do not have the required {\Lya} narrow-band imaging
to investigate this issue.
}.
This factor is used throughout the paper, except in
Figs.~\ref{fig:contfluxden_compare} and~\ref{fig:afterglow_vs_host}.

\begin{figure*}
\makebox[\textwidth]{\includegraphics[width=1.00\textwidth,bb=10 573 564 745]{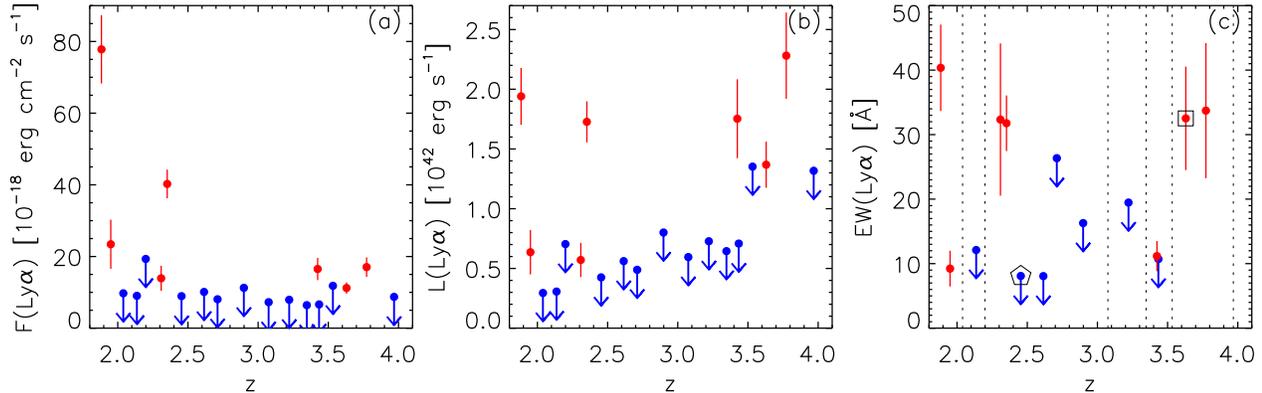}}
\caption{
{\Lya} fluxes, luminosities and (rest-frame) EWs 
for the 20 hosts in the sample.
The dotted lines in panel~(c) represent hosts for which no limit could be
placed on the EW, due to detecting neither {\Lya}
nor the continuum in the spectra.
The two systems where a neighboring object was fitted and subtracted
are marked: GRB\,070721B (box) and GRB\,070802 (pentagon).
The symbol colors simply reflect the symbol type:
red = detections, blue = upper limits.
\label{fig:Lya_flux_lum_EW_vs_z}
}
\end{figure*}

\begin{figure*}
\makebox[\textwidth]{\includegraphics[width=1.00\textwidth,bb=31 272 547 751]{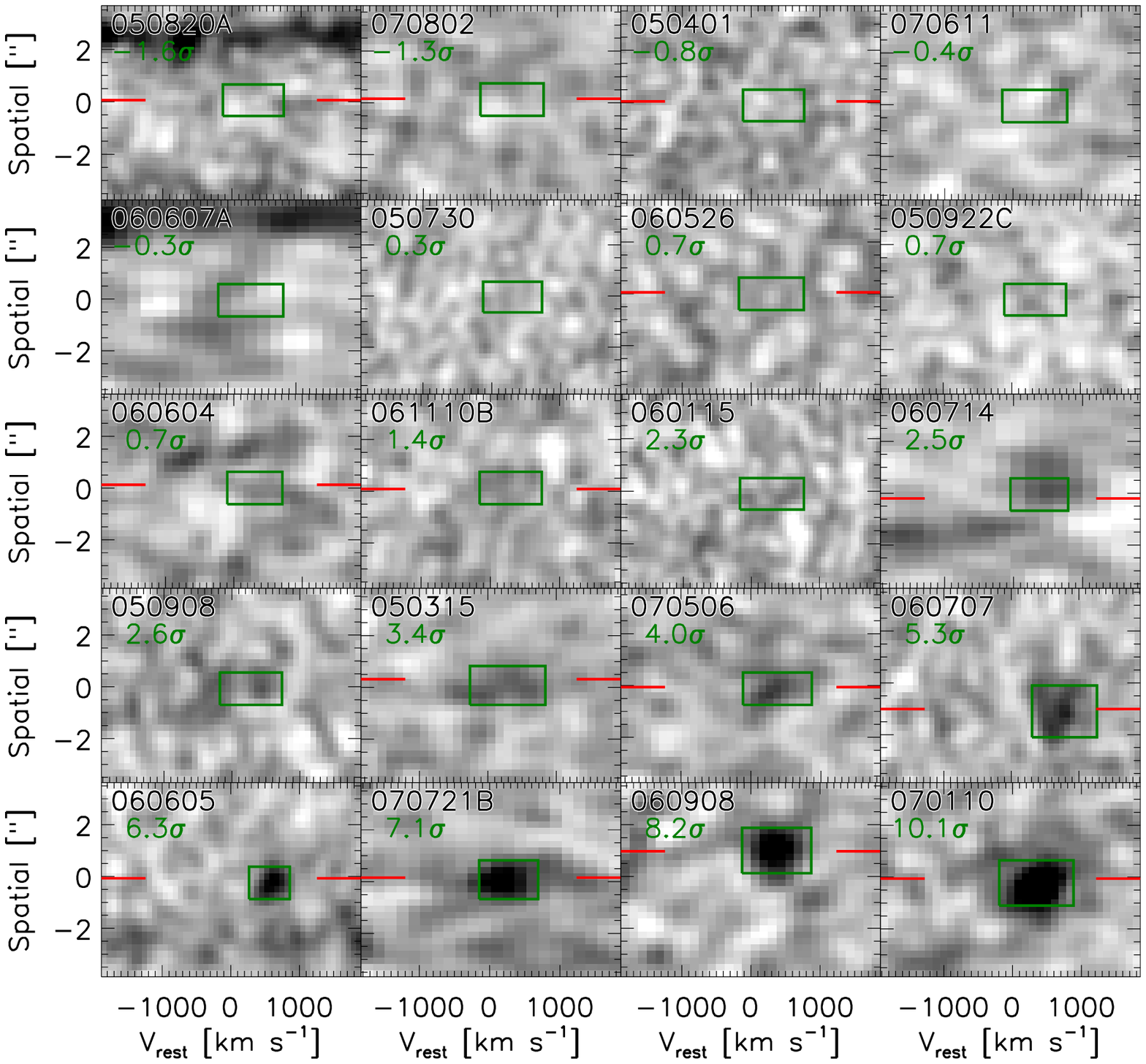}}
\caption[]{
The obtained 2D spectra, centered on where {\Lya} is expected.
The zero point for the rest-frame velocity is defined by the afterglow
redshift listed in Table~\ref{tab:sample_obs}.
The zero point for the spatial (angular) position corresponds to
the afterglow position.
The shown sections have size
$3800\,\mathrm{km}\,\mathrm{s}^{-1} \times 7.5''$.
The spectra have been smoothed by a Gaussian with FWHM = 3\,px.
The intensity cuts are the same for all the panels, in units of the
noise in the given spectrum, allowing a visual comparison of the
significance of the features in the different panels.
The green rectangle marks the aperture within which the {\Lya} flux and its
uncertainty are measured; the aperture centers and widths are listed in
Table~\ref{tab:Lya_flux}.
On the panels is stated by how many sigma {\Lya} is detected;
the panels are sorted by this.
The red horizontal lines indicate where the continuum
(if detected at $\ge3\sigma$) is located,
as defined by the spatial centroid in the aperture in which the
continuum is measured, cf.\ Fig.~\ref{fig:spatialprofile_montage_sample}.
\label{fig:2Dspec_montage_sample}
}
\end{figure*}

\begin{figure*}
\makebox[\textwidth]{\includegraphics[width=1.00\textwidth,bb=20 272 547 751]{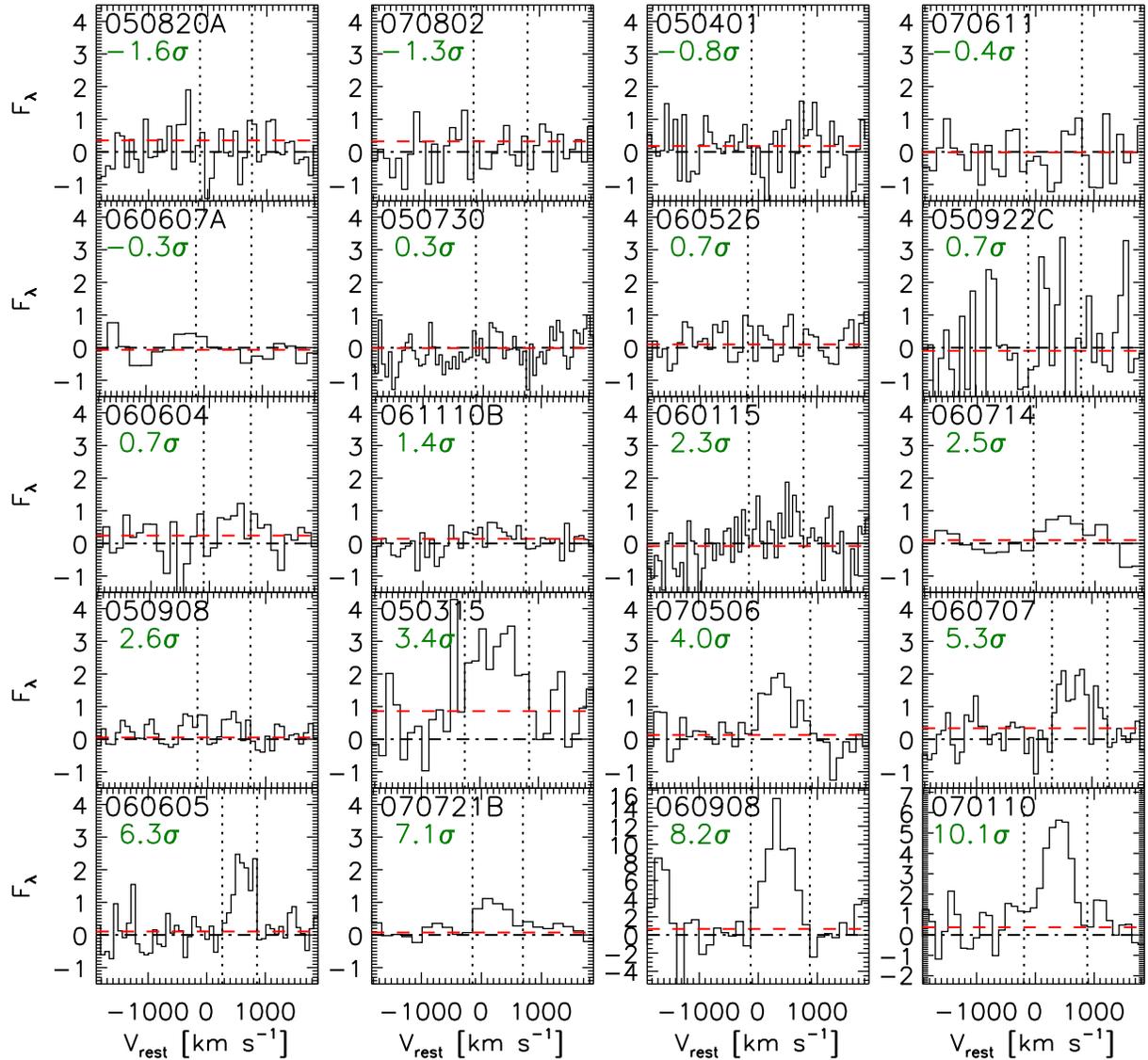}}
\caption[]{
1D spectra, derived by a straight sum over the spatial aperture
(cf.\ Table~\ref{tab:Lya_flux}).
The flux densities $F_\lambda$ are in units of
$10^{-18}\,\mathrm{erg}\,\mathrm{s}^{-1}\,\mathrm{cm}^{-2}\,\mathrm{\AA}^{-1}$.
No smoothing has been applied.
The detection significance of the {\Lya} emission line is
given on the panel;
the panels are sorted by this.
The vertical dotted lines mark the velocity limits of the
{\Lya} aperture, cf.\ Table~\ref{tab:Lya_flux}.
The horizontal dot-dashed (black) line simply indicates zero,
while the horizontal dashed (red) line indicates the continuum,
measured as the mean level in a 175\,{\AA} rest-frame wide window
on the red side of {\Lya}.
The plotted range in velocity corresponds to the range
of the 2D spectra shown in Fig.~\ref{fig:2Dspec_montage_sample}.
\label{fig:1Dspec_montage_sample}
}
\end{figure*}

\begin{figure*}
\makebox[\textwidth]{\includegraphics[width=1.00\textwidth,bb=34 272 547 751]{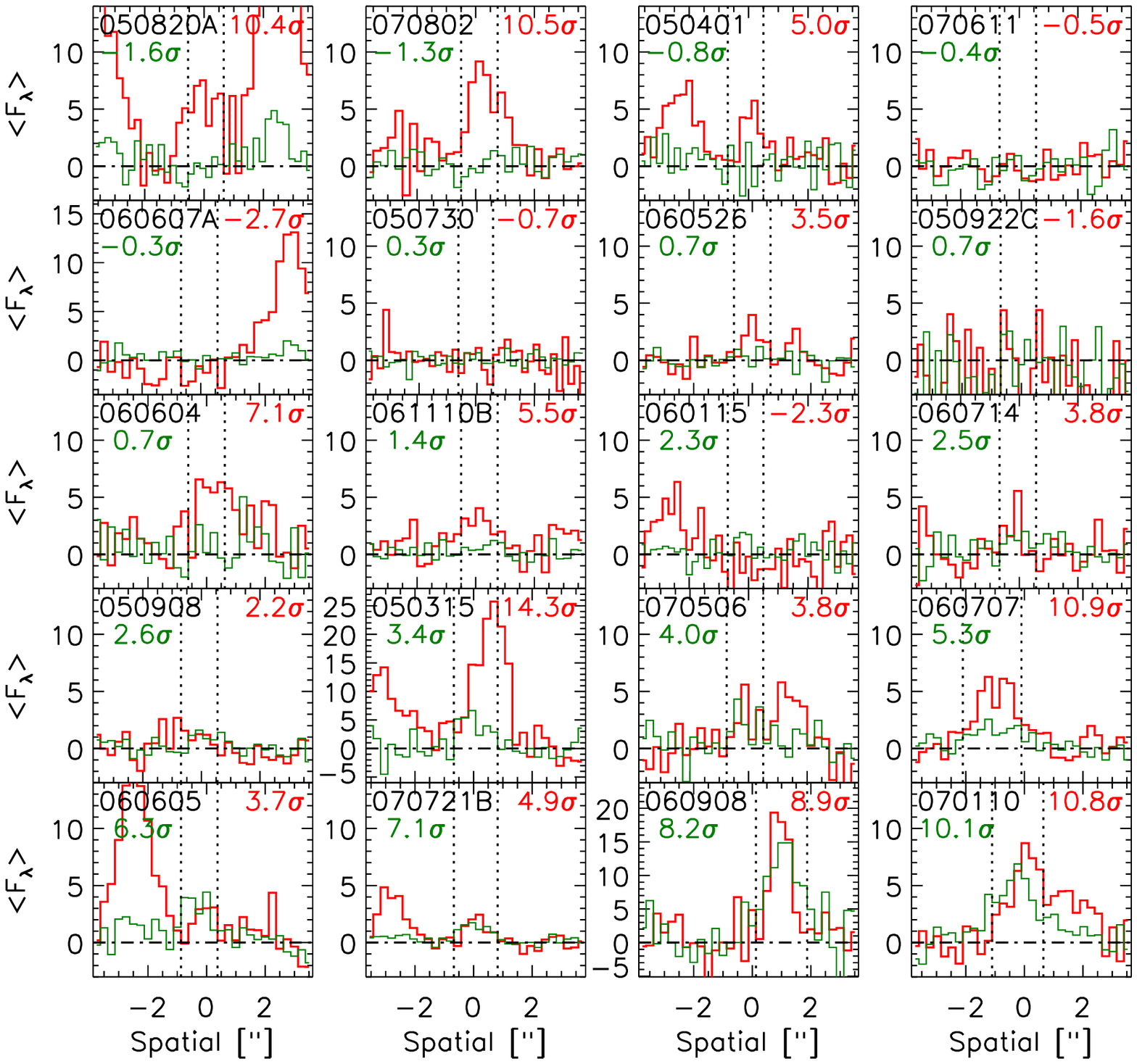}}
\caption[]{
Spatial profiles, i.e.\
the flux densities averaged over a given wavelength range
versus spatial coordinate in the 2D spectrum.
Red thick curve: $\langle F_\lambda \rangle$ calculated over
the 175$\,${\AA} rest-frame wide continuum window
on the red side of {\Lya}, in units of
$10^{-20}\,\mathrm{erg}\,\mathrm{s}^{-1}\,\mathrm{cm}^{-2}\,\mathrm{\AA}^{-1}$.
Green curve: $\langle F_\lambda \rangle$ calculated over
the wavelength range of the {\Lya} aperture, in units of
$10^{-19}\,\mathrm{erg}\,\mathrm{s}^{-1}\,\mathrm{cm}^{-2}\,\mathrm{\AA}^{-1}$.
No smoothing has been applied.
The detection significance of the continuum in the above-mentioned red window is
given on the right hand side of the panel in red, and
the detection significance of the {\Lya} emission line is
given on the left hand side of the panel in green.
The vertical dotted lines mark the spatial limits of the
{\Lya} aperture, cf.\ Table~\ref{tab:Lya_flux}.
The plotted range in spatial coordinate corresponds to the range
of the 2D spectra shown in Fig.~\ref{fig:2Dspec_montage_sample}.
The panels are sorted by {\Lya} detection significance.
\label{fig:spatialprofile_montage_sample}
}
\end{figure*}

The {\Lya} fluxes were measured from the 2D spectra using
a rectangular aperture defined in terms of
$v$, the rest-frame velocity with respect to the afterglow redshift
(cf.\ Table~\ref{tab:sample_obs}), and
$s$, the spatial offset along the slit with respect
to the afterglow position. 
If {\Lya} was detected, the aperture was centered on the line and
the width of the aperture was adjusted accordingly (as illustrated in
the last 7 panels of
Figs.~\ref{fig:2Dspec_montage_sample}--\ref{fig:spatialprofile_montage_sample}).
If {\Lya} was not detected, an aperture of width
$900\,\mathrm{km}\,\mathrm{s}^{-1} \times 1.2''$ was used,
centered at $300\,\mathrm{km}\,\mathrm{s}^{-1}$ in $v$
(a value typical for the detected lines, cf.\ \S\ref{sec:lya_velocity_offset})
and at $s = 0.0''$.
The aperture was defined in terms of integer pixels for simplicity.
The centers and widths of the apertures are listed in Table~\ref{tab:Lya_flux}.
The width in $v$ can be compared to the spectral
resolution expressed as a rest-frame velocity, $c/R$,
which is $\la 700\,\mathrm{km}\,\mathrm{s}^{-1}$,
cf.\ Table~\ref{tab:wlrange_specres}.

The fluxes within the {\Lya} apertures in the 2D spectra were integrated
by summing the flux densities
(in $\mathrm{erg}\,\mathrm{s}^{-1}\,\mathrm{cm}^{-2}\,\mathrm{\AA}^{-1}$)
and multiplying by the spectral pixel size (in {\AA})\@.
Uncertainties were calculated by propagating the individual uncertainties
from the 2D sigma spectra (\S\ref{sec:datareduc}). 
The continuum contribution was subtracted (if positive)
and the errors propagated, to give
the continuum-subtracted {\Lya} fluxes listed in Table~\ref{tab:Lya_flux}.
If the {\Lya} emission was not detected at 3$\sigma$ (with the 
$\sigma$ being the uncertainty on the continuum-subtracted {\Lya} flux),
we give the 3$\sigma$ upper limit in the table.
{\Lya} emission was detected in 7 hosts out of 20:
GRB\,050315,
GRB\,060605,
GRB\,060707,
GRB\,060908,
GRB\,070110,
GRB\,070506, and
GRB\,070721B\@.
Of these, only one was known to be a {\Lya} emitter from
the literature (GRB\,070110, \citealt{Fynbo_etal:2009}).
{\Lya} emission was not detected at 3$\sigma$ for GRB\,060714,
only at 2.5$\sigma$, but {\Lya} emission was convincingly detected in
the afterglow spectrum, cf.\ \S\ref{sec:lya_afterglow_host_comparison}
below.
Note that the 20 hosts
in Figs.~\ref{fig:2Dspec_montage_sample}--\ref{fig:spatialprofile_montage_sample}
are sorted by {\Lya} detection significance, as listed on the panels.

Rest-frame EWs of {\Lya} were calculated from the
{\Lya} fluxes and from the continuum flux densities
measured in the spectra in a 175\,{\AA} [rest-frame] wide window
on the red side of Lya centered at 1309\,{\AA} (cf.\ above),
and the errors propagated. 
Throughout the remainder of this paper it is implicit that the EWs are 
rest-frame values.
The results are summarized in Fig.~\ref{fig:Lya_flux_lum_EW_vs_z}
which shows the {\Lya} fluxes, luminosities and EWs
(or upper limits in case of non-detections) versus redshift.

\begin{deluxetable}{lrrrrccccc}
\tabletypesize{\small}
\tablewidth{0pt}
\tablecaption{{\Lya} measurements from the spectra\label{tab:Lya_flux}}
\tablehead{
\colhead{Name} &
\multicolumn{4}{c}{\hrulefill\ {\Lya} aperture \hrulefill} &
\colhead{$F$({\Lya})} &
\colhead{$L$({\Lya})} &
\colhead{$F_\lambda$(cont.)} &
\colhead{EW({\Lya})} &
\colhead{$v$({\Lya})} \\
\colhead{} &
\colhead{c($v$)} &
\colhead{c($s$)} &
\colhead{w($v$)} &
\colhead{w($s$)} &
\colhead{} &
\colhead{} &
\colhead{} &
\colhead{} &
\colhead{} \\
\colhead{(1)} &
\colhead{(2)} &
\colhead{(3)} &
\colhead{(4)} &
\colhead{(5)} &
\colhead{(6)} &
\colhead{(7)} &
\colhead{(8)} &
\colhead{(9)} &
\colhead{(10)}
}
\startdata
GRB\,050315  & 290 & $ 0.08$ & 1102 &  1.50 & $ 23.4 \pm  6.8 $ & $ 0.64 \pm 0.19 $ & $ 86.0 \pm  6.0 $                  & $  9.2 \pm  2.8 $                  & $  283 \pm 62 $ \\
GRB\,050401  & 326 & $-0.10$ &  888 &  1.20 & $<11.2          $ & $<0.80          $ & $ 18.0 \pm  3.6 $                  & $<16.3          $                  &       \nodata   \\
GRB\,050730  & 302 & $-0.04$ &  856 &  1.20 & $< 8.7          $ & $<1.32          $ & $< 8.0          $                  &   \nodata                          &       \nodata   \\
GRB\,050820A & 309 & $ 0.06$ &  878 &  1.20 & $<10.1          $ & $<0.56          $ & $ 35.1 \pm  3.4 $                  & $< 8.1          $                  &       \nodata   \\
GRB\,050908  & 281 & $-0.08$ &  914 &  1.25 & $< 6.4          $ & $<0.64          $ & $< 7.0          $                  &   \nodata                          &       \nodata   \\
GRB\,050922C & 342 & $-0.10$ &  902 &  1.20 & $<19.3          $ & $<0.70          $ & $<18.6          $                  &   \nodata                          &       \nodata   \\
GRB\,060115  & 303 & $-0.10$ &  938 &  1.20 & $<11.8          $ & $<1.35          $ & $<10.5          $                  &   \nodata                          &       \nodata   \\
GRB\,060526  & 294 & $ 0.00$ &  941 &  1.25 & $< 7.9          $ & $<0.73          $ & $  9.8 \pm  2.8 $                  & $<19.5          $                  &       \nodata   \\
GRB\,060604  & 307 & $ 0.00$ &  806 &  1.25 & $< 9.0          $ & $<0.31          $ & $ 23.9 \pm  3.3 $                  & $<12.1          $                  &       \nodata   \\
GRB\,060605  & 564 & $-0.25$ &  594 &  1.25 & $ 17.0 \pm  2.7 $ & $ 2.28 \pm 0.36 $ & $ 10.6 \pm  2.8 $                  & $ 33.7 \pm 10.5 $                  & $  620 \pm 26 $ \\
GRB\,060607A & 337 & $-0.05$ &  953 &  1.25 & $< 7.3          $ & $<0.60          $ & $< 7.3          $                  &   \nodata                          &       \nodata   \\
GRB\,060707  & 788 & $-1.10$ &  961 &  2.00 & $ 16.5 \pm  3.1 $ & $ 1.75 \pm 0.33 $ & $ 33.4 \pm  3.1 $                  & $ 11.2 \pm  2.3 $                  & $  742 \pm 38 $ \\
GRB\,060714  & 373 & $-0.05$ &  837 &  1.25 & $< 8.1          $ & $<0.49          $ & $  9.9 \pm  2.6 $                  & $<26.3          $                  &       \nodata   \\
GRB\,060908  & 372 & $ 1.00$ & 1002 &  1.75 & $ 77.8 \pm  9.5 $ & $ 1.94 \pm 0.24 $ & $ 66.8 \pm  7.5 $                  & $ 40.4 \pm  6.7 $                  & $  347 \pm 31 $ \\
GRB\,061110B & 294 & $ 0.12$ &  896 &  1.25 & $< 6.6          $ & $<0.71          $ & $ 14.4 \pm  2.6 $                  & $<10.7          $                  &       \nodata   \\
GRB\,070110  & 345 & $-0.22$ & 1078 &  1.75 & $ 40.2 \pm  4.0 $ & $ 1.73 \pm 0.17 $ & $ 37.8 \pm  3.5 $                  & $ 31.8 \pm  4.3 $                  & $  358 \pm 26 $ \\
GRB\,070506  & 379 & $-0.12$ &  983 &  1.25 & $ 13.9 \pm  3.5 $ & $ 0.57 \pm 0.14 $ & $ 13.0 \pm  3.5 $                  & $ 32.3 \pm 11.8 $                  & $  360 \pm 62 $ \\
GRB\,070611  & 319 & $-0.05$ &  951 &  1.25 & $< 9.7          $ & $<0.29          $ & $< 8.7          $                  &   \nodata                          &       \nodata   \\
GRB\,070721B & 271 & $ 0.08$ &  839 &  1.50 & $ 11.2 \pm  1.6 $ & $ 1.37 \pm 0.19 $ & $  7.4 \pm  1.5 $\tablenotemark{a} & $ 32.5 \pm  8.0 $\tablenotemark{a} & $  212 \pm 31 $ \\
GRB\,070802  & 321 & $ 0.12$ &  941 &  1.25 & $< 8.9          $ & $<0.43          $ & $ 32.2 \pm  3.1 $\tablenotemark{a} & $< 8.1          $\tablenotemark{a} &       \nodata   \\
\enddata
\tablenotetext{a}{The listed error reflects the random error only.
A systematic error due to the fitting and subtraction of a
neighboring object is likely present.}
\tablecomments{
Columns (2)--(5) define the aperture in the 2D spectrum
within which the {\Lya} flux was measured.
Specifically, the columns give the center (c) and width (w) of the aperture in terms of
$v$ (in $\mathrm{km}\,\mathrm{s}^{-1}$), the rest-frame velocity with respect to
the afterglow redshift (cf.\ Table~\ref{tab:sample_obs}), and
$s$ (in arcsec), the spatial offset along the slit with respect to the afterglow position.
The subsequent columns are:
(6)~$F$({\Lya}), the {\Lya} emission line flux, in units of
$10^{-18}\,\mathrm{erg}\,\mathrm{cm}^{-2}\,\mathrm{s}^{-1}$;
(7)~$L$({\Lya}), the {\Lya} luminosity, in units of
$10^{42}\,\mathrm{erg}\,\mathrm{s}^{-1}$;
(8)~$F_\lambda$(cont.), the continuum flux density measured in a 175\,{\AA}
rest-frame wide aperture (centered at 1309\,{\AA}) on the red side of {\Lya}, in units of
$10^{-20}\,\mathrm{erg}\,\mathrm{cm}^{-2}\,\mathrm{s}^{-1}\,\mathrm{\AA}^{-1}$;
(9)~EW({\Lya}), the rest-frame {\Lya} equivalent width, in {\AA};
(10)~$v$({\Lya}), the rest-frame velocity of the {\Lya} emission line
with respect to the afterglow redshift,
i.e.\ the centroid of $v$ within the aperture, in $\mathrm{km}\,\mathrm{s}^{-1}$.
The listed error on $v$({\Lya}) is only based on the errors on the fluxes within
the aperture, and does not include the error on the afterglow redshift.
All upper limits are 3$\sigma$.
The flux calibrated science spectra and their uncertainties (sigma spectra)
have been multiplied by a factor of 2.0 
to correct for slit loss and aperture loss (\S\ref{sec:lya_fluxes_ews}).
This has made all the numbers in columns (6), (7) and (8)
be larger by this factor.
}
\end{deluxetable}

The spectra of the 20 systems are illustrated in three figures:
(i)~Smoothed 2D spectra centered on the {\Lya} lines are shown in Fig.~\ref{fig:2Dspec_montage_sample}
with the used aperture indicated;
(ii)~1D spectra (not smoothed) are shown in Fig.~\ref{fig:1Dspec_montage_sample}; and
(iii)~Spatial profiles (also not smoothed) are shown in Fig.~\ref{fig:spatialprofile_montage_sample}.

\subsection{Velocity offset between {\boldLya} emission and low-ionization
absorption lines in the afterglow spectra}
\label{sec:lya_velocity_offset}

\begin{figure}[htbp]
\includegraphics[scale=1.00,bb=9 545 229 751]{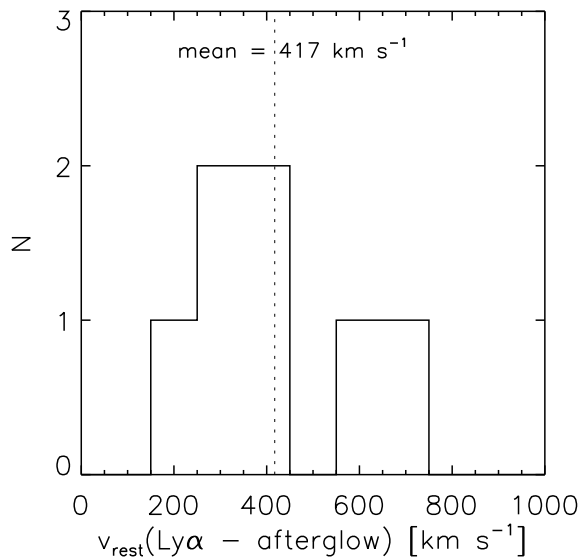}
\caption{Distribution of velocity offsets between the centroid of the
{\Lya} emission in the GRB host spectrum and
low-ionization interstellar absorption lines in the GRB afterglow spectrum.
\label{fig:vel_hist}
}
\end{figure}

For the 7 hosts with a $\ge3\sigma$ detection of {\Lya} emission,
the centroid in $v$ within the {\Lya} aperture was measured
(see Table~\ref{tab:Lya_flux}).
Since the afterglow redshift defines the zero point of $v$,
this velocity measurement is the
velocity offset between the {\Lya} emission centroid and low-ionization 
interstellar absorption in the GRB afterglow spectrum.
A histogram of this offset for the 7 hosts is given in Fig.~\ref{fig:vel_hist}.
The range is 200--700$\,\mathrm{km}\,\mathrm{s}^{-1}$,
consistent with the few values for GRB hosts reported in the literature
(cf.\ Table~\ref{tab:known_Lya_hosts}).
The histogram resembles the distribution of 
velocity offsets between {\Lya} emission and low-ionization
interstellar absorption in LBGs
(\citealt{Adelberger_etal:2003,Shapley_etal:2003}; see also \citealt{Pettini_etal:2000}).
The distribution is also in agreement with the
velocity offsets for two {\Lya}-selected galaxies reported by
\citet{McLinden_etal:2011} using the [\ion{O}{3}] emission line
to define the systemic velocity.
It should be noted that the assumption that the afterglow redshift
provides the systemic velocity is only valid on average (over a sample
of hosts). For individual hosts the GRB sightline may probe a region
of the host that has a non-zero velocity due to the internal kinematics
of the galaxy (e.g.\ rotation).
This cannot be a large effect, since otherwise the measured velocity offsets
of {\Lya} (Fig.~\ref{fig:vel_hist}) would not all have the same sign.

It should be noted that what we measure is simply the centroid of the
{\Lya} emission line in our GRB host spectra, which may not be identical
to the peak of the line if the line is asymmetrical.
Our spectra (Fig.~\ref{fig:1Dspec_montage_sample})
do not have sufficient spectral resolution or S/N
to investigate this issue.

The origin of the offset is most likely a combination of radiative transfer 
of the resonantly scattered {\Lya} photons and a star-formation driven outflow 
from the host galaxy (for a full discussion of these effects we refer to 
\citealt{Fynbo_etal:2010}, their \S4.3).
It should be emphasized that the velocities presented
in Table~\ref{tab:Lya_flux} and Fig.~\ref{fig:vel_hist}
simply represent the centroid of the {\Lya} line in the spectrum
(with respect to the afterglow redshift);
they do not directly translate into outflow velocities.
In the outflow scenario several factors affect the observed redwards shift
of the {\Lya} velocity centroid with respect to the systemic velocity.
High column densities and (to a lesser extent) low temperatures push the
{\Lya} peak further from the systemic velocity \citep{Harrington:1973}.
The velocity shift also increases with increasing outflow velocities,
up to $\sim10^3\,\mathrm{km}\,\mathrm{s}^{-1}$, where the peak starts to
drift back toward the systemic velocity \citep[e.g.][]{Verhamme_etal:2006}.
On the other hand, if dust is present, preferentially the wings of the line
are removed, effectively reducing the shift \citep{Laursen_etal:2009b}.
This effect is stronger the more homogeneous the medium is, since clumpiness
of the gas and dust facilitates the escape of {\Lya} photons
\citep{Neufeld:1991,Hansen_Oh:2006}.

\subsection{{\boldLya} emission comparison: host and afterglow spectra}
\label{sec:lya_afterglow_host_comparison}

\begin{figure}
\includegraphics[scale=0.75,bb=9 409 294 751,clip=]{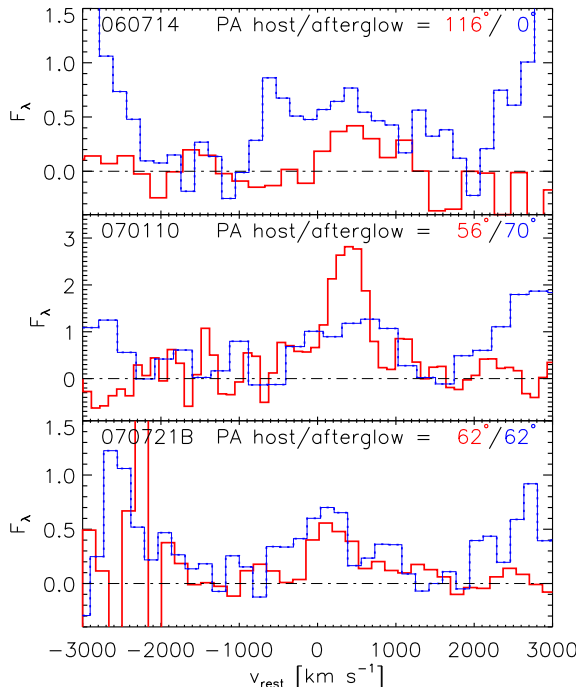}
\caption[]{
Comparison of host spectra (red solid lines) with
afterglow spectra (blue dotted+solid lines).
The three GRBs shown are those 
for which {\Lya} emission was detected 
in the afterglow spectrum
superimposed on the damped {\Lya} absorption trough.
The position angle (PA) of the slit is given on the panels;
note the large PA difference between host and afterglow spectroscopy
for GRB\,060714.
The spectra have been corrected for Galactic extinction.
$F_\lambda$ is given in units of 
$10^{-18}\,\mathrm{erg}\,\mathrm{cm}^{-2}\,\mathrm{s}^{-1}\,\mathrm{\AA}^{-1}$.
In the comparison for GRB\,070721B it should be noted that only
the host spectrum has had a neighboring galaxy subtracted (cf.\ \S\ref{sec:lya_fluxes_ews}),
a procedure that decreases the overall flux level by about 0.1 in the plotted units.
The afterglow spectra were taken from
\citet{Jakobsson_etal:2006} [GRB\,060714] and
\citet{Fynbo_etal:2009} [GRB\,070110 and GRB\,070721B].
\label{fig:afterglow_vs_host}
None of the spectra in this figure have been corrected for
slit losses or aperture losses.
}
\end{figure}

For three of the hosts in the sample
{\Lya} emission was detected directly in the afterglow spectra:
GRB\,060714 \citep{Jakobsson_etal:2006},
GRB\,070110 \citep{Fynbo_etal:2009}, and also marginally in
GRB\,070721B \citep{Fynbo_etal:2009}.
Figure~\ref{fig:afterglow_vs_host} compares the spectra.
For GRB\,070110 and GRB\,070721B the {\Lya} flux measured
from the afterglow and host spectroscopy
is consistent within the errors (at 2 sigma).
However, for GRB\,060714 there is a significant difference,
with the {\Lya} flux
from the afterglow spectrum being almost a factor of 4 larger than in the 
host spectrum reported here. In the afterglow spectrum the {\Lya} emission
appears to be very extended both spatially and in velocity space. In addition,
the position angles of the two observations are nearly perpendicular (0$^\circ$
east of north for the afterglow spectrum and 116$^\circ$ for the host galaxy
spectrum). Hence, we suspect the cause of the difference is a higher
slit loss in the host galaxy spectrum.
It is possible that other of the 12 hosts where we do not detect
{\Lya} emission at $\ge3\sigma$ actually would have been detected
as {\Lya} emitters had we used a slit at a different position angle
or a wider slit.

\section{Discussion and summary}
\label{sec:summary_discussion}

In this work we have carried out a systematic search for {\Lya} photons from
GRB host galaxies selected from the larger well-defined TOUGH sample
of such galaxies
presented in \citet{Hjorth_etal:2012}. Unlike previous studies
(cf.\ \citealt{Fynbo_etal:2003a}) we find that
{\Lya} emission is not ubiquitous among GRB host galaxies.
Of the 20 host galaxies studied here we detect (at $3\sigma$) {\Lya} emission
from 7 of them (with the {\Lya} [rest-frame] EW in the range 9--40\,{\AA}),
derive $3\sigma$ upper limits on the {\Lya} EW for 7 of them
(in the range 8--26\,{\AA}),
while we obtain no constraints on the {\Lya} EW for the last 6 hosts
(due to neither detecting the continuum nor {\Lya} emission in the
spectra, both at $3\sigma$), cf.\ Table~\ref{tab:Lya_flux}.
Out of the 14 hosts with either a {\Lya} EW or an upper limit on the EW,
8 hosts have {\Lya} EW less than 20\,{\AA} (rest-frame), which is the
typical limit in narrow-band surveys for {\Lya} emitters.
For the 7 detections, the measured EWs in the range 9--40\,{\AA} are low
compared to the distribution of EWs found for narrow-band selected galaxies at
similar redshifts \citep{Gronwall_etal:2007,Grove_etal:2009,Nilsson_etal:2009a}.

The {\Lya} luminosities for the 7 GRB hosts with detected {\Lya} emission are in the range
(0.6--2.3)$\,\times 10^{42}\,\mathrm{erg}\,\mathrm{s}^{-1}$.            
Such fairly low {\Lya} luminosities are only probed by a few studies
of {\Lya} emitting galaxies,
e.g.\ \citet{Rauch_etal:2008,Grove_etal:2009,Cassata_etal:2011}.
The {\Lya} luminosity can be translated into a star-formation rate (SFR),
assuming no dust extinction,
as
\begin{equation}
\mathrm{SFR} = \frac{ L(\mbox{\Lya}) }
                    { 1.1 \times 10^{42}\,\mathrm{erg}\,\mathrm{s}^{-1} } \,
                    M_\sun\,\mathrm{yr}^{-1} \enspace ,
\label{eq:SFR}
\end{equation}
using the relation between SFR and $L$(H$\alpha$) 
from \citet{Kennicutt:1998a} and the case B recombination ratio
$L$({\Lya})/$L$(H$\alpha$) = 8.7 \citep{Brocklehurst:1971}.
The observed range in {\Lya} luminosity for the 7 detections
would translate into a range in SFR of 0.6--2.1$\,M_\sun\,\mathrm{yr}^{-1}$,  
but the assumption of no dust extinction is likely not always correct,
as illustrated by the trends of {\Lya} luminosity and EW with
afterglow spectral index discussed in \S\ref{sec:betaOX}.
If dust is present then Eq.~(\ref{eq:SFR}) provides a lower limit
of the SFR\@.

\subsection{Comparison with LBGs}
\label{sec:LBGs}

\begin{figure}
\includegraphics[scale=0.90,bb=10 190 248 750]{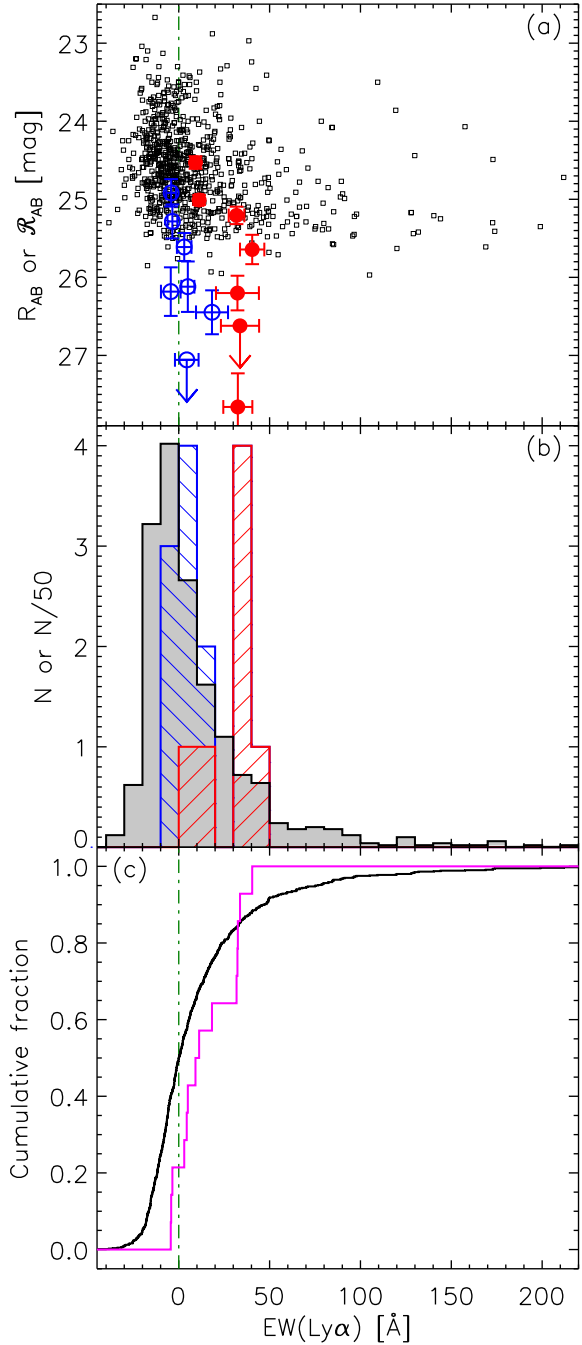}
\caption{
Comparison of {\Lya} EWs between the LBGs from \citet{Shapley_etal:2003} and
the GRB host galaxies from this work.
Small black points and grey/black histogram: LBGs ($N=803$).
Other points: GRB host galaxies ($N=14$), in (a) and (b) split into
galaxies with detected {\Lya} emission (red filled circles and histogram)
and without (blue open circles and histogram).
\label{fig:Lya_EW_comparison}
}
\end{figure}

Lyman-break selection and GRB selection are complementary methods to identify
samples of galaxies at high redshift. In this section and in
Fig.~\ref{fig:Lya_EW_comparison} we carry out a comparison of
the LBGs from \citet{Shapley_etal:2003} with
the GRB host galaxies from this work.

The 20 GRB host galaxies from this work fall into 3 categories.
For 6 galaxies we detected neither the continuum nor
{\Lya} emission in the spectra, and these galaxies are omitted from
the analysis.
For 7 galaxies we detected both the continuum and {\Lya} emission in
the spectra, and for these we use the {\Lya} EWs reported in
Table~\ref{tab:Lya_flux}.
For another 7 galaxies we detected the continuum but not {\Lya} emission in
the spectra. In Table~\ref{tab:Lya_flux} and the rest of the paper we have
reported the $3\sigma$ upper limits on the {\Lya} EWs.
An example is GRB\,061110B where the measured [rest-frame] EW is
$5.0 \pm 3.6$\,{\AA}, which we replaced by the $3\sigma$ upper limit of
EW $<$ 10.7\,{\AA}\@.
This procedure implicitly assumes that {\Lya} can only be in emission.
However, the LBGs from \citet{Shapley_etal:2003} often show significant
{\Lya} \emph{absorption} (negative EWs), so in order to make a fair
comparison with that sample, we use the measured EWs also for the
7 GRB host galaxies without detected {\Lya} emission\footnote{%
These EWs are:
GRB\,050401:  $ -4.4 \pm 5.4$\,{\AA};
GRB\,050820A: $ -4.2 \pm 2.7$\,{\AA};
GRB\,060526:  $  4.4 \pm 6.5$\,{\AA};
GRB\,060604:  $  3.0 \pm 4.0$\,{\AA};
GRB\,060714:  $ 18.3 \pm 8.8$\,{\AA};
GRB\,061110B: $  5.0 \pm 3.6$\,{\AA}; and
GRB\,070802:  $ -3.5 \pm 2.7$\,{\AA}.
}.

Figure~\ref{fig:Lya_EW_comparison}(a)
shows apparent $R$-band magnitude\footnote{%
The apparent magnitudes in Fig.~\ref{fig:Lya_EW_comparison}(a)
are on the AB system and have been corrected for Galactic extinction.
The filter used for the LBGs from \citet{Shapley_etal:2003} is $\mathcal{R}$,
see \citet{Steidel_Hamilton:1993}.
}
versus {\Lya} EW\@.
The 803 LBGs are shown as small black open squares.
The 7 GRB host galaxies with {\Lya} emission detected at $3\sigma$ are shown
as red filled circles, and
the 7 GRB host galaxies without such detected {\Lya} emission are shown
as blue open circles.
The plot 
shows that the GRB hosts from this work typically are fainter than
the LBGs from \citet{Shapley_etal:2003}.
This is also the case for the luminosities,
since the redshift distributions of the two samples are fairly similar:
LBGs:              $\langle z \rangle = 3.0$, sd = 0.3;
GRB host galaxies: $\langle z \rangle = 2.8$, sd = 0.6 (with sd being the standard deviation).
It is tempting to define a faint subset of the LBG sample that is
better matched to our sample, but \citet{Shapley_etal:2003} conclude
that the redshift incompleteness at fainter magnitudes
\citep[say fainter than $\mathcal{R} \approx 24.5$, cf.\ Fig.~7 in][]{Shapley_etal:2003}
is likely such that preferentially galaxies without (strong) {\Lya} emission
are missing.
This might argue for only comparing our GRB hosts with a \emph{bright}
LBG subsample, but then the luminosity difference would be substantial.
We will therefore simply use the full \citet{Shapley_etal:2003} for
comparison with our GRB host sample.

Figure~\ref{fig:Lya_EW_comparison}(b) shows histograms of the EWs:
grey filled histogram: LBGs;
hacthed histograms: GRB hosts galaxies, with blue and red having the same
meaning as in panel~(a).
The LBG histogram has been scaled down by a factor of 50, but is
otherwise identical to Fig.~8 in \citet{Shapley_etal:2003}.

Figure~\ref{fig:Lya_EW_comparison}(c) shows the cumulative EW distributions:
smooth black curve: LBGs; jagged magenta curve: the 14 GRB host galaxies.
A Kolmogorov--Smirnov (K--S) test \citep[e.g.][]{Press_etal:1992}
gives a 1.7\% probability that the two samples are drawn from the same
parent distribution.
In other words, we detect a difference at 98.3\% confidence
between the two samples.
This result is driven by the lack of GRB host galaxies with
substantial {\Lya} absorption (i.e.\ with EWs below $-5$\,{\AA})\@.

\citet{Fynbo_etal:2003a} also found a significant difference
(99.8\% confidence) between the EWs of 5 pre-{\swift} GRB host galaxies and
an approximation of the \citet{Shapley_etal:2003} distribution.
If we use our updated compilation of the EWs for these 5 pre-{\swift} hosts
and carry out
a K--S test against the \citet{Shapley_etal:2003} sample, we get a similar
result, namely a difference that is significant at 99.2\% confidence.
We compare the pre-{\swift} sample with the sample from this work in
\S\ref{sec:betaOX}.

\subsection{The relation between afterglow spectral index and host {\boldLya} emission, and comparison with pre-{\swift} studies}
\label{sec:betaOX}

Remarkably, substantially larger EWs were found in the previous,
pre-{\swift} studies of {\Lya} emission from GRB hosts despite the fact that
these studies targeted a much smaller sample
\citep[][and references therein]{Fynbo_etal:2003a}.
Our updated compilation of the EWs for the 5 pre-{\swift} hosts
studied by \citet{Fynbo_etal:2003a}
is given in the first 5 rows of Table~\ref{tab:known_Lya_hosts};
the 3 large values around 70--100\,{\AA} are noteworthy.
A K--S test comparing the EWs of the pre-{\swift} sample ($N=5$)
with the sample from this work ($N=14$, cf.\ \S\ref{sec:LBGs})
gives an 8\% probability that the two samples are drawn from the same
parent distribution.
This is marginal evidence for a difference.
This difference could therefore be a 
chance effect, but another plausible explanation is 
different biases in the two samples.

\begin{figure*}
\makebox[\textwidth]{
  \includegraphics[scale=0.40,bb = 18 179 592 690]{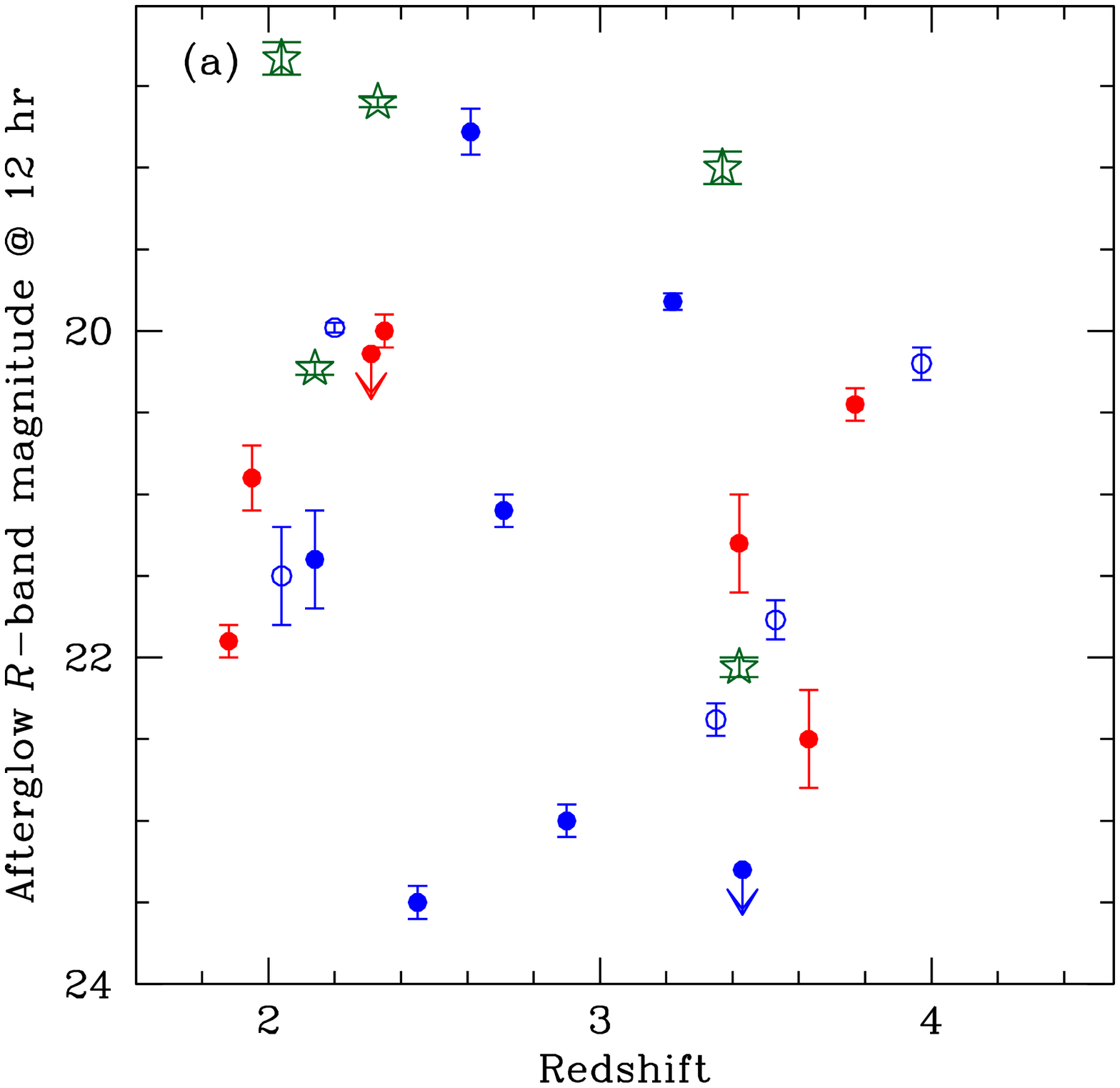} 
  \includegraphics[scale=0.40,bb = 18 179 592 690]{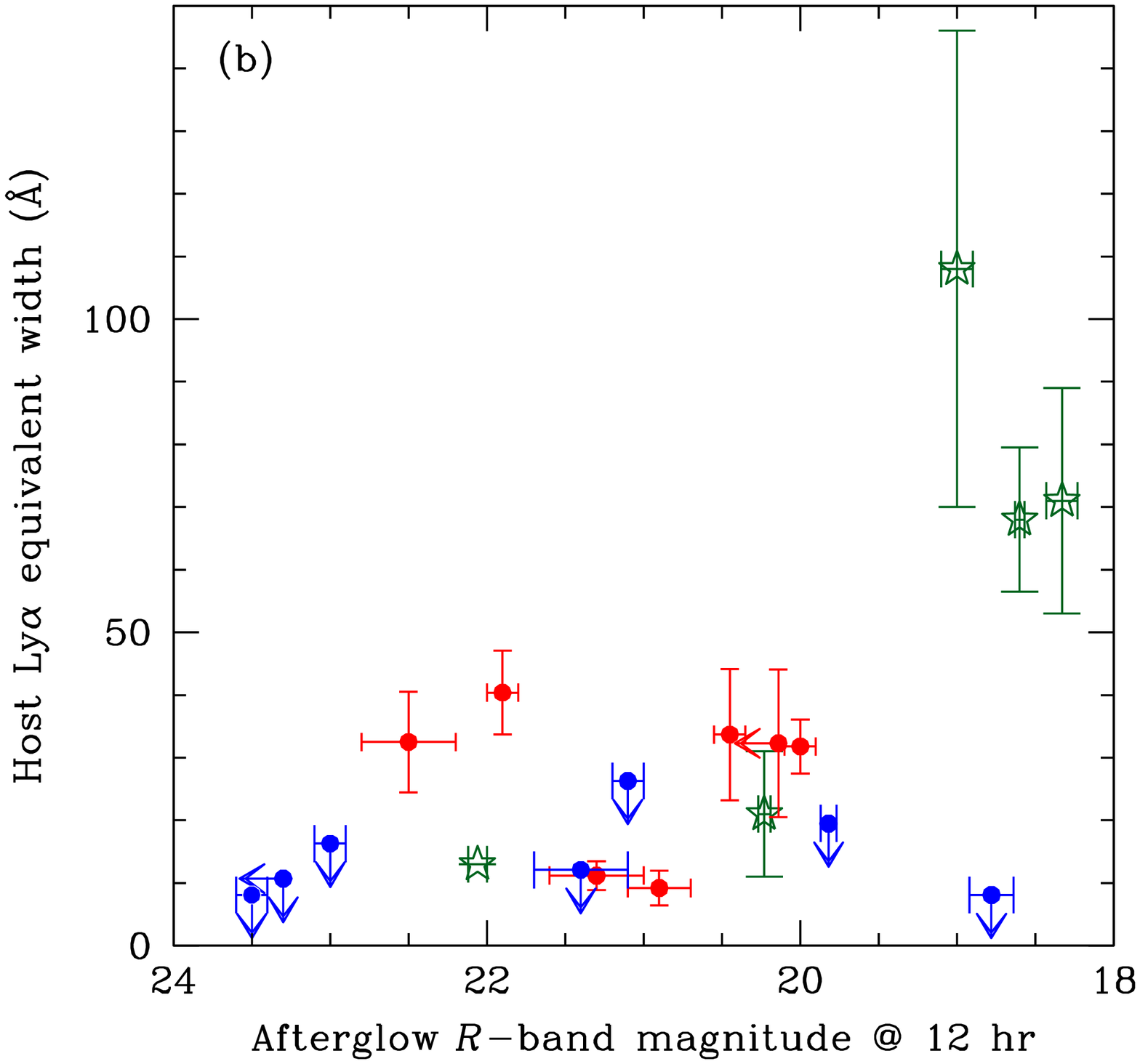} 
}
\caption{
Afterglow $R$--band magnitude at 12$\,$hr after the burst,
versus redshift (panel~a) and {\Lya} EW (panel~b).
Open green stars: the pre-{\swift} sample of 5 GRB host {\Lya} emitters.
Other symbols: the sample of 20 hosts from this work:
red filled circles: hosts with a {\Lya} detection,
blue filled cirles: hosts without a {\Lya} detection but with an upper limit on the {\Lya} EW,
blue open circles (panel~a only): hosts with no contraints on the {\Lya} EW\@.
\label{fig:afterglow_mag}
}
\end{figure*}

The present sample is based on an underlying X--ray selected sample
of 69 bursts
\citep[the TOUGH sample, see \S\ref{sec:parent_sample} and][]{Hjorth_etal:2012}
which is nearly unbiased.
The sample of 20 bursts followed up for {\Lya} spectroscopy in this work
(i.e.\ those with a known afterglow redshift in the range 1.8--4.5)
is biased since an optical afterglow was \emph{de facto}
required, and since some bursts in the TOUGH sample were without
a determined redshift at the time of the target selection for
the {\Lya} spectroscopy and thus could be in the targeted
redshift range of 1.8--4.5 (indeed, 7 of these bursts were recently
found to be at $z$ = 1.8--4.5 from X--shooter host spectroscopy,
see \citealt{Kruhler_etal:2012:TOUGH5} and below,
while 20 of the TOUGH bursts still do not have a determined redshift,
cf.\ \S\ref{sec:targetsel}).
The pre-{\swift} sample of 5 bursts 
\citep[][and references therein]{Fynbo_etal:2003a}
is even more biased towards relatively bright optical afterglows
due to the larger times to localize the burst and larger
localization uncertainties
(see also \citealt{Kann_etal:2010}).
This is shown in Fig.~\ref{fig:afterglow_mag}(a), where we plot the
afterglow $R$--band magnitude at 12$\,$hr after the burst
(see Table~\ref{tab:afterglow_mag}) versus redshift
for the pre-{\swift} sample (open green stars)
and the {\swift} sample from this work (other symbols).
In panel~(b) we plot EW({\Lya}) (detections, and for our sample
also upper limits) versus afterglow magnitude.
Comparing the two samples suggests that the larger
{\Lya} EWs for the pre-{\swift} sample is related to brighter afterglows,
which in turn could be related to galaxies having less dust.
On the other hand, within the sample from this work there is no evidence for a correlation
between {\Lya} EW and afterglow magnitude. 

\begin{deluxetable}{llcrl}
\tabletypesize{\small}
\tablewidth{0pt}
\tablecaption{Afterglow $R$--band magnitudes at 12$\,$hr after the burst --- The 5 pre-{\swift} {\Lya} emitters and the 20 hosts from this work\label{tab:afterglow_mag}}
\tablehead{
\colhead{Name} &
\colhead{EW} &
\colhead{$R$} &
\colhead{Ref.} &
\colhead{Comment}
}
\startdata
GRB\,971214  & yes     & $ 22.06 \pm 0.06$ &     (1) & \\                                                                            
GRB\,000926  & yes     & $ 18.33 \pm 0.10$ &     (2) & Extrapolated from $R=19.33$ at 20.7\,hr using $\alpha=1.69$ \\                
GRB\,011211  & yes     & $ 20.23 \pm 0.04$ &     (3) & \\                                                                            
GRB\,021004  & yes     & $ 18.60 \pm 0.03$ &     (4) & \\                                                                            
GRB\,030323  & yes     & $ 19.00 \pm 0.1~$ &     (5) & \\                                                                            
\hline                                                                                                                               
GRB\,050315  & yes     & $ 20.90 \pm 0.20$ &     (6) & \\                                                                            
GRB\,050401  & UL      & $ 23.00 \pm 0.10$ &     (7) & \\                                                                            
GRB\,050730  & \nodata & $ 20.20 \pm 0.1~$ &     (8) & \\                                                                            
GRB\,050820A & UL      & $ 18.78 \pm 0.14$ &     (6) & \\                                                                            
GRB\,050908  & \nodata & $ 22.38 \pm 0.10$ &     (9) & Data at 0.68, 4.25 and 22.12\,hr, $\alpha=0.78$ \\                            
GRB\,050922C & \nodata & $ 19.98 \pm 0.03$ &     (6) & \\                                                                            
GRB\,060115  & \nodata & $ 21.77 \pm 0.12$ &    (10) & \\                                                                            
GRB\,060526  & UL      & $ 19.82 \pm 0.05$ &    (11) & \\                                                                            
GRB\,060604  & UL      & $ 21.40 \pm 0.30$ &    (12) & \\                                                                            
GRB\,060605  & yes     & $ 20.45 \pm 0.1~$ &    (13) & \\                                                                            
GRB\,060607A & \nodata &    \nodata        & \nodata & Unusual light curve --- unable to interpolate \\                              
GRB\,060707  & yes     & $ 21.30 \pm 0.30$ &    (14) & Interpolation between two datapoints \\                                       
GRB\,060714  & UL      & $ 21.10 \pm 0.1~$ &    (15) & \\                                                                            
GRB\,060908  & yes     & $ 21.90 \pm 0.1~$ &    (16) & \\                                                                            
GRB\,061110B & UL      & $>23.30         $ &    (17) & Limit assuming $\alpha=0.5$ from the VLT observation at 2.3\,hr \\            
GRB\,070110  & yes     & $ 20.00 \pm 0.1~$ &    (18) & Converted from $V$--band \\                                                   
GRB\,070506  & yes     & $>20.14         $ &    (19) & Limit assuming $\alpha=0.5$ from the VLT observation at 3.6\,hr \\            
GRB\,070611  & \nodata & $ 21.50 \pm 0.30$ &    (17) & Extrapolated from $R=21.0$ at 7.7\,hr adopting $\alpha = 1.0\pm0.5$ \\        
GRB\,070721B & yes     & $ 22.50 \pm 0.30$ &    (20) & Extrapolated from $R=23.0\pm0.1$ at 17.6\,hr adopting $\alpha = 1.0\pm0.5$ \\ 
GRB\,070802  & UL      & $ 23.50 \pm 0.1~$ &    (21) & \\                                                                            
\enddata
\tablecomments{
The first 5 bursts are from the pre-{\swift} sample
\citep[][and references therein]{Fynbo_etal:2003a}, and the remaining 20 bursts
are from the {\swift}-based sample studied in this work.
The EW column indicates what type of information is available about
the {\Lya} EW of the host (detection, upper limit or no constraint); the actual values
are given in Tables~\ref{tab:Lya_flux} and \ref{tab:known_Lya_hosts}.
$R$ is the afterglow $R$--band magnitude at 12$\,$hr after the burst.
In the cases where we directly read the afterglow magnitude at 12$\,$hr after the burst
from a plot in the stated reference we have assigned a magnitude error of 0.1.
The parameter $\alpha$ is the slope of the light curve: $F(t) \propto t^{-\alpha}$.
}
\tablerefs{
 (1)~\citet{Diercks_etal:1998};
 (2)~\citet{Fynbo_etal:2001a};
 (3)~\citet{Jakobsson_etal:2004:GRB011211};
 (4)~\citet{Holland_etal:2003};
 (5)~\citet{Vreeswijk_etal:2004};
 (6)~\citet{Kann_etal:2010};
 (7)~\citet{Watson_etal:2006};
 (8)~\citet{Pandey_etal:2006};
 (9)~Our measurement using archival data from ESO program 275.D-5022 (PI: Chincarini);
(10)~Our measurement using archival data from ESO program 076.A-0392 (PI: Tagliaferri);
(11)~\citet{Thone_etal:2010};
(12)~\citet{Tanvir_etal:2006};
(13)~\citet{Ferrero_etal:2009};
(14)~\citet{deUgartePostigo_etal:2006:GCN5290} and \citet{Jakobsson_etal:2006:GCN5298};
(15)~\citet{Krimm_etal:2007};
(16)~\citet{Covino_etal:2010};
(17)~\citet{Fynbo_etal:2009};
(18)~\citet{Troja_etal:2007};
(19)~\citet{Fynbo_etal:2009} but corrected for typo: correct value is $R=19.5$ at 3.6\,hr;
(20)~Our measurement using data from our own ESO program 079.D-0429 (PI: Vreeswijk);
(21)~\citet{Kruhler_etal:2008}.
}

\end{deluxetable}

To further examine the role of dust we turn to the afterglow spectral index
$\betaOX$, defined by
\begin{equation}
\betaOX = \frac{ \log\left[F_\nu(\nu_\mathrm{opt})/F_\nu(\nu_\mathrm{X})\right] }
               { \log\left[\nu_\mathrm{X}/\nu_\mathrm{opt}\right] } \enspace ,
\label{eq:betaOX}
\end{equation}
where $F_\nu$ is the flux density of the afterglow and
where $\nu_\mathrm{opt}$ and $\nu_\mathrm{X}$ are representative center
frequencies (pivotal frequencies) of the optical and X--ray bands, respectively.
If $F_\nu$ were a single power-law between $\nu_\mathrm{opt}$ and $\nu_\mathrm{X}$
it would have the form $F_\nu \propto \nu^{-\betaOX}$.
A low value of $\betaOX$ indicates suppression of the optical emission
compared to the X--ray flux. For low-redshift events
(e.g.\ for $z \le 4.5$ as considered here), where the optical
is not cut off by inter-galactic medium 
absorption, $\betaOX$ is thus connected to dust
extinction along the GRB sightline (e.g.\ \citealt{Fynbo_etal:2009}).
In particular, assuming standard synchrotron theory, $\betaOX$ cannot be
intrinsically smaller than 0.5, and therefore bursts with $\betaOX < 0.5$ are
referred to as ``dark bursts'' \citep{Jakobsson_etal:2004b}, although moderate
extinction can be present also in bursts with larger values of $\betaOX$.

\begin{deluxetable}{lccrcccccccc}
\tabletypesize{\small}
\tablewidth{0pt}
\tablecaption{Known GRB host {\Lya} emitters: afterglow spectral slope and host {\Lya} properties\label{tab:known_Lya_hosts}}
\tablehead{
\colhead{Name} &
\colhead{$z$} &
\colhead{T} &
\colhead{$\betaOX$} &
\colhead{Ref.} &
\colhead{EW({\Lya})} &
\colhead{Ref.} &
\colhead{$F$({\Lya})} &
\colhead{Ref.} &
\colhead{$L$({\Lya})} &
\colhead{$v$({\Lya})} &
\colhead{Ref.}
}
\startdata
GRB\,971214  & 3.42 &  no &    0.64 &     (1) &                    $ 13          $ &     (6) &    $ 6.2 \pm 0.7 $                  &    (13) & $  0.66 \pm 0.07 $ &   \nodata       & \nodata \\
GRB\,000926  & 2.04 &  no &    0.87 &     (1) &                      $ 71 \pm 18 $ &     (7) &    $ 149 \pm  11 $\tablenotemark{b} &     (7) & $  4.51 \pm 0.33 $ &   \nodata       & \nodata \\
GRB\,011211  & 2.14 &  no &    0.98 &     (1) &                      $ 21 \pm 10 $ &     (8) &  $ 33.6 \pm  9.6 $\tablenotemark{b} &     (8) & $  1.14 \pm 0.33 $ &   \nodata       & \nodata \\
GRB\,021004  & 2.33 &  no &    0.93 &     (1) &                    $ 68 \pm 11.5 $ &     (9) &    $ 313 \pm  64 $\tablenotemark{b} &    (14) & $ 13.14 \pm 2.67 $ & $ 530         $ &    (14) \\
GRB\,030323  & 3.37 &  no & \nodata & \nodata &                     $ 108 \pm 38 $ &    (10) &       $ 12 \pm 1 $                  &    (15) & $  1.23 \pm 0.10 $ & $ 151 \pm  46 $ &    (15) \\
GRB\,050315  & 1.95 & yes &    0.63 &     (2) & $  9.2 \pm  2.8 $                  &    (11) &  $ 23.4 \pm  6.8 $                  &    (11) &  $ 0.64 \pm 0.19 $ & $  283 \pm 62 $ &    (11) \\
GRB\,060605  & 3.77 & yes &    1.00 &     (2) & $ 33.7 \pm 10.5 $                  &    (11) &  $ 17.0 \pm  2.7 $                  &    (11) &  $ 2.28 \pm 0.36 $ & $  620 \pm 26 $ &    (11) \\
GRB\,060707  & 3.42 & yes &    0.73 &     (2) & $ 11.2 \pm  2.3 $                  &    (11) &  $ 16.5 \pm  3.1 $                  &    (11) &  $ 1.75 \pm 0.33 $ & $  742 \pm 38 $ &    (11) \\
GRB\,060714  & 2.71 & yes &    0.77 &     (2) &                            \nodata & \nodata &  $ 17.3          $\tablenotemark{b} &    (16) & $  1.05          $ &   \nodata       & \nodata \\
GRB\,060908  & 1.88 & yes &    0.80 &     (3) & $ 40.4 \pm  6.7 $                  &    (11) &  $ 77.8 \pm  9.5 $                  &    (11) &  $ 1.94 \pm 0.24 $ & $  347 \pm 31 $ &    (11) \\
GRB\,060926  & 3.21 &  no &    0.87 &     (4) &                            \nodata & \nodata &  $ 62.1 \pm  4.9 $\tablenotemark{b} &    (17) & $  5.65 \pm 0.45 $ & $ 311         $ &    (17) \\
GRB\,061222A & 2.09 &  no & $<$0.22 &     (2) &                    $ 31          $ &    (12) &   $ 168          $\tablenotemark{b} &    (18) & $  5.39          $ &   \nodata       & \nodata \\
GRB\,070110  & 2.35 & yes &    0.77 &     (2) & $ 31.8 \pm  4.3 $                  &    (11) &  $ 40.2 \pm  4.0 $                  &    (11) &  $ 1.73 \pm 0.17 $ & $  358 \pm 26 $ &    (11) \\
GRB\,070506  & 2.31 & yes &    0.93 &     (2) & $ 32.3 \pm 11.8 $                  &    (11) &  $ 13.9 \pm  3.5 $                  &    (11) &  $ 0.57 \pm 0.14 $ & $  360 \pm 62 $ &    (11) \\
GRB\,070721B & 3.63 & yes &    0.72 &     (2) & $ 32.5 \pm  8.0 $\tablenotemark{a} &    (11) &  $ 11.2 \pm  1.6 $                  &    (11) &  $ 1.37 \pm 0.19 $ & $  212 \pm 31 $ &    (11) \\
GRB\,071031  & 2.69 &  no &    0.97 &     (2) &                            \nodata & \nodata &  $ 23.6 \pm  2.7 $\tablenotemark{b} &    (17) & $  1.41 \pm 0.16 $ & $ 254         $ &    (17) \\
GRB\,090205  & 4.65 &  no &    0.98 &     (5) &                            \nodata & \nodata &  $ 23.6 \pm  4.9 $\tablenotemark{b} &    (19) & $  5.17 \pm 1.08 $ & $ 180 \pm 153 $ &    (19) \\
\enddata
\tablenotetext{a}{The listed error reflects the random error only.
A systematic error due to the fitting and subtraction of a
neighboring object is likely present.}
\tablenotetext{b}{The published {\Lya} flux or the provided spectrum was not corrected for Galactic extinction, but we have applied the correction.}
\tablecomments{
{T indicates whether the host is part of the TOUGH sample \citep[\S\ref{sec:parent_sample};][]{Hjorth_etal:2012} studied in this work.}
$\betaOX$ is the afterglow optical-to-X--ray spectral slope (see Eq.~\ref{eq:betaOX}),
where optical means $R$--band (unless otherwise stated) and X--ray means 3 keV\@.
EW({\Lya}) is the rest-frame {\Lya} emission line EW, in {\AA}\@.
$F$({\Lya}) is the {\Lya} emission line flux, in units of $10^{-18}\,\mathrm{erg}\,\mathrm{cm}^{-2}\,\mathrm{s}^{-1}$.
$L$({\Lya}) is the {\Lya} emission line luminosity, in units of $10^{42}\,\mathrm{erg}\,\mathrm{s}^{-1}$.
$v$({\Lya}) is the rest-frame velocity centroid of the {\Lya} emission line with respect to the afterglow redshift, in $\mathrm{km}\,\mathrm{s}^{-1}$.
``Ref." gives the reference for the preceeding column.
The bursts up to and including GRB\,030323 are pre-{\swift}, while the remaining bursts are from {\swift}.
All {\Lya} fluxes and luminosities are corrected for Galactic extinction.
Note that GRB\,030429 is not included, since even though its spectrum showed
an indication of {\Lya} emission, it was not statistically significant
($\la 2\sigma$) \citep{Jakobsson_etal:2004a}.
}
\tablerefs{
(1)~\citet{Jakobsson_etal:2004b};
(2)~\citet{Fynbo_etal:2009};
(3)~Our calculation, using $R$--band data from \citet{Covino_etal:2010} and X--ray data from \citet{Evans_etal:2007,Evans_etal:2009} (we note that the $\betaOX$ value in \citealt{Fynbo_etal:2009} is in error);
(4)~Our calculation, using archival $R$--band data from ESO program 077.D-0805 (PI: Tagliaferri) and X--ray data from \citet{Evans_etal:2007,Evans_etal:2009};
(5)~Our calculation, using $I$--band photometry from \citet{DAvanzo_etal:2010} (rather than $R$--band, due to the high redshift) and X--ray data from \citet{Evans_etal:2007,Evans_etal:2009};
(6)~Our measurement, using the host spectrum from \citet{Kulkarni_etal:1998};
(7)~\citet{Fynbo_etal:2002}, with $F$({\Lya}) corrected for continuum contribution as prescribed in that paper;
(8)~\citet{Fynbo_etal:2003a};
(9)~\citet{Jakobsson_etal:2005};
(10)~Our calculation, using the line flux from \citet{Vreeswijk_etal:2004} and deriving the continuum flux density from the F606W photometry from \citet{Wainwright_etal:2007} (correcting for {\Lya} forest absorption and imposing a slitloss; these effects almost cancel out);
(11)~This work;
(12)~Our measurement, using the host spectrum from \citet{Perley_etal:2009};
(13)~\citet{Kulkarni_etal:1998};
(14)~\citet{Moller_etal:2002};
(15)~\citet{Vreeswijk_etal:2004};
(16)~\citet{Jakobsson_etal:2006};
(17)~Our measurement, using the afterglow spectrum from \citet{Fynbo_etal:2009};
(18)~D. Perley, priv.\ comm., cf.\ \citet{Perley_etal:2009};
(19)~\citet{DAvanzo_etal:2010}.
\vspace*{-3.0em} 
}
\end{deluxetable}

\begin{figure*}
\makebox[\textwidth]{\includegraphics[scale=0.99,bb=10 522 471 743]{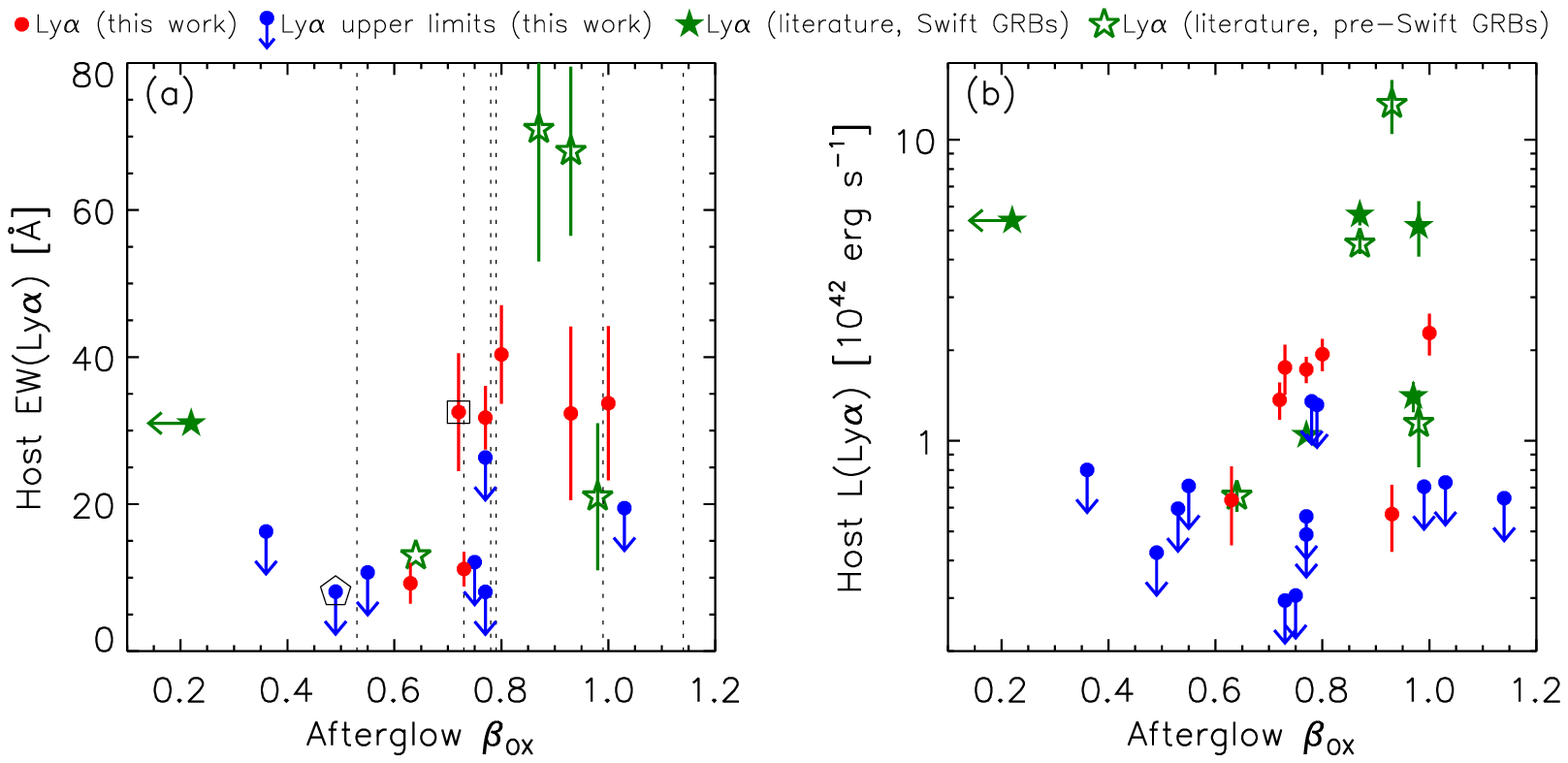}} 
\caption{
Host {\Lya} EW (panel a) and host {\Lya} luminosity (panel b) versus afterglow spectral index $\betaOX$.
The GRBs in the sample of this paper are shown as
red filled circles ({\Lya} detections),
blue upper limits ({\Lya} non-detections), and
dotted lines (unconstrained {\Lya} EWs, panel~a only).
Additional GRB hosts with {\Lya} detections from the literature
are shown as green stars (open stars: pre-{\swift}, filled stars: {\swift}).
These bursts are not in the sample of this paper,
except for GRB\,060714: this burst has a
{\Lya} detection from the literature,
whereas our data resulted in an upper limit,
as discussed in \S\ref{sec:lya_afterglow_host_comparison}.
This burst is therefore plotted twice (panel~b only):
as a detection and as an upper limit.
Some of the literature bursts plotted in panel~(b)
are absent from panel~(a) due to not having a measured EW,
which in turn is due to only an afterglow spectrum being available.
The plotted data for all the {\Lya} detections
(both from this work and from the literature)
are given in Table~\ref{tab:known_Lya_hosts}.
The non-detections (this work only) are given
in Table~\ref{tab:Lya_flux}, with $\betaOX$ taken from
\citet{Fynbo_etal:2009}.
Note that GRB\,030323
(listed in Table~\ref{tab:known_Lya_hosts})
is not plotted here due to not having X--ray observations and hence no $\betaOX$.
The two systems where a neighboring object was fitted and subtracted
are marked: GRB\,070721B (box) and GRB\,070802 (pentagon).
Note the logarithmic axis for the {\Lya} luminosity.
\label{fig:Lya_vs_betaOX}
}
\end{figure*}

We have compiled a list of all known GRB hosts with {\Lya}
emission, both from this work and from the literature, including both
pre-{\swift} and {\swift} bursts, see Table~\ref{tab:known_Lya_hosts}.
The table lists both the afterglow $\betaOX$ and the host {\Lya}
emission properties.
We have corrected the literature {\Lya} fluxes for Galactic extinction
where needed.
The table only contains hosts with a {\Lya} detection.
The {\Lya} upper limits from this work are in Table~\ref{tab:Lya_flux}
(with $\betaOX$ available for all bursts from \citealt{Fynbo_etal:2009}).

In Fig.~\ref{fig:Lya_vs_betaOX} we plot host EW({\Lya}) and host $L$({\Lya})
versus afterglow $\betaOX$,
both for the hosts from this work and for the additional
hosts with {\Lya} detections from the literature.
The plots show a lack of hosts with high EW({\Lya}) and large $L$({\Lya})
at the low end of the $\betaOX$ range.
This indicates that dust extinction is important in reducing the strength
of the {\Lya} line.
GRB\,061222A goes against the trend,
with a very low $\betaOX$ (namely an upper limit\footnote{%
The upper limit comes from the afterglow only having an $R$--band upper limit.
The afterglow has a $K_\mathrm{s}$--band detection
\citep{Cenko_Fox:2006},
which gives $\betaOX = 0.10$ (with ``O'' now signifying $K_\mathrm{s}$--band rather
than optical/$R$--band), which is even more constraining.
}
of $\betaOX < 0.22$) and detected {\Lya} emission.
This can be explained by the fact that $\betaOX$ only probes the
afterglow sightline, whereas the host {\Lya} emission is a quantity that is
global for the galaxy.
There are indeed several cases where a dark GRB exploded in an overall
blue galaxy, e.g.\
GRB\,070306 \citep{Jaunsen_etal:2008}
and
GRB\,100621A \citep{Kruhler_etal:2011}.
It is also seen from  Fig.~\ref{fig:Lya_vs_betaOX} that
the hosts of the pre-{\swift} GRBs with high EW({\Lya}) and large $L$({\Lya})
are preferentially found
at the high end of the $\betaOX$ range. This suggests that the 
pre-{\swift} sample
(shown as open green stars in Fig.~\ref{fig:Lya_vs_betaOX})
discussed by \citet{Fynbo_etal:2003a}
is more biased against dusty sightlines.

Our finding that {\Lya} emission is not ubiquitous among GRB host
galaxies has implications on how well GRBs trace the overall massive
star-formation activity and on the nature of GRB progenitors. Given
that {\Lya} photons are more easily destroyed by dust than other
UV photons due to resonant scattering, it has been argued that GRB
hosts have low dust content. This could be due, among other reasons,
to low metallicity, in agreement with the prediction of the collapsar
model \citep{MacFadyen_Woosley:1999,Yoon_Langer:2005,Woosley_Heger:2006}.
Our analysis of a larger sample of GRB hosts shows that
{\Lya} emission is less ubiquitous than previously found based on
a much smaller sample \citep{Fynbo_etal:2003a,Jakobsson_etal:2005},
so that the above argument is not valid. Whereas other mechanisms than dust
can reduce the strength of the {\Lya} line (e.g.\ the geometry of the
interstellar medium), the trend visible in Fig.~\ref{fig:Lya_vs_betaOX}
suggests that the strength of the {\Lya} line is related to the presence
of dust. We note that the objects in the sample studied in this work all
have a redshift measured from the optical afterglow, hence they are biased
against very dusty systems. If the connection between the presence of dust
and the weakness of the {\Lya} line holds, we expect that the hosts of
optically-obscured (i.e.\ dark) GRBs should have even less prominent
{\Lya} emission.

The recent work of \citet{Kruhler_etal:2012:TOUGH5}
provides additional insight. VLT/X-shooter was used to target
several TOUGH hosts that lacked redshifts. For 7 of the TOUGH
hosts the found redshift was in the range $z$ = 1.8--4.5,
and these are thus hosts missed by the  target selection
for this work (cf.\ \S\ref{sec:targetsel}).
The redshifts were based on detecting one or more of the
following emission lines: [\ion{O}{2}], H$\beta$, [\ion{O}{3}] and H$\alpha$.
In no cases was {\Lya} detected.
These 7 bursts mostly have low $\betaOX$ values\footnote{%
The $\betaOX$ values are:
GRB\,050819:    $<$0.90,
GRB\,050915A:   $<$0.44,
GRB\,051001:    $<$0.56,
GRB\,060814: $<$$-$0.06,
GRB\,070103:    $<$0.48,
GRB\,070129:       0.62, and
GRB\,070419B:      0.25
\citep[][and references therein]{Fynbo_etal:2009}.
}.
While the lack of {\Lya} emission still has to be quantified
in terms of upper limits on the EWs,
the \citet{Kruhler_etal:2012:TOUGH5} result 
supports the picture that weak or absent {\Lya} emission
is at least in part caused by dust.

\acknowledgments

The Dark Cosmology Centre is funded by the Danish National Research
Foundation.
BMJ and JPUF acknowledge support from the ERC-StG grant EGGS-278202.
This research has made use of the NASA/IPAC Extragalactic Database (NED)
which is operated by the Jet Propulsion Laboratory, California Institute
of Technology, under contract with the National Aeronautics and Space
Administration.
This work made use of data supplied by the UK Swift Science Data Centre
at the University of Leicester.
We thank S.~George~Djorgovski for providing the spectrum of the GRB\,971214 host
and
Dan~Perley for providing the spectrum of the GRB\,061222A host,
cf.\ Table~\ref{tab:known_Lya_hosts}.
Peter~Laursen is thanked for discussions about {\Lya} radiative transfer.
Alice~E.~Shapley is thanked for discussions and for
providing the data on the LBGs from
\citet{Shapley_etal:2003} in machine-readable format.
Thomas~Kr{\"u}hler is thanked for discussions about the X-shooter
host spectroscopy.
We thank the anonymous referee for comments that helped improve the
presentation.

{\it Facilities:} \facility{VLT (FORS1)}.

\bibliography{papers_cited_by_milvang,papers_tmp}

\appendix

\section{Observations obtained of systems not in the TOUGH sample}
\label{appendix}

\begin{deluxetable}{lllrllccc}
\tabletypesize{\small}
\tablewidth{0pt}
\tablecaption{Observed systems not in the TOUGH sample\label{tab:sample_obs_nonsample}}
\tablehead{
\colhead{Name} &
\colhead{$z$} &
\colhead{Ref.} &
\colhead{$R_\mathrm{host}$} &
\colhead{Grism+filter} &
\colhead{CCD} &
\colhead{$T_\mathrm{exp}^\mathrm{total}$} &
\colhead{Seeing} &
\colhead{$A_V$} \\
\colhead{} &
\colhead{} &
\colhead{} &
\colhead{(mag)} &
\colhead{} &
\colhead{} &
\colhead{(hr)} &
\colhead{(arcsec)} &
\colhead{(mag)}
}
\startdata
 \object{GRB 050603}  & N/A\tablenotemark{a}    & (1)     & $   >26.6$                  & 600B         & old &   2.2  & $<1.1 $ & 0.092 \\ 
 \object{GRB 060223A} & 4.406                   & (2)     & $   >26.3$                  & 600R+GG435   & old &   2.1  & $ 0.7 $ & 0.389 \\ 
 \object{GRB 070810A} & 2.17                    & (3)     & $   >26.7$                  & 600B         & new &   1.5  & $ 0.7 $ & 0.072 \\ 
\enddata
\tablenotetext{a}{The redshift $z=2.821$ for GRB\,050603 from
\citet{Berger_Becker:2005} is likely wrong: it was derived based on a reported 
very bright emission line interpreted as {\Lya} in the afterglow spectrum,
but in our deep host spectrum we do not detect any {\Lya} emission;
we derive a $3\sigma$ upper limit on the {\Lya} flux at $z=2.821$ of
$4.7 \times 10^{-18}\,\mathrm{erg}\,\mathrm{s}^{-1}\,\mathrm{cm}^{-2}$
(cf.\ Table~\ref{tab:Lya_flux_nonsample}).
}
\tablecomments{
See Table~\ref{tab:sample_obs} for further information.
}
\tablerefs{
(1)~\citet{Berger_Becker:2005};
(2)~\citet{Chary_etal:2007};
(3)~\citet{Thone_etal:2007:GCN_6741}.
}
\end{deluxetable}

Three systems which were not in the final TOUGH sample
were also observed, see Table~\ref{tab:sample_obs_nonsample}.
The reason for these 3 systems not being in the TOUGH sample are as follows:
GRB\,050603 and GRB\,060223A did not have an XRT position distributed
within 12 hours (although an XRT observation had been made within
12 hours), and
GRB\,070810A did not have a Sun distance greater than 55$^\circ$
(its Sun distance was 49$^\circ$).
The TOUGH sample criteria are described in \citet{Hjorth_etal:2012}
and are summarised in \S\ref{sec:parent_sample}.

\begin{figure*}[htbp]
\makebox[\textwidth]{\includegraphics[width=1.00\textwidth,bb=31 625 547 751]{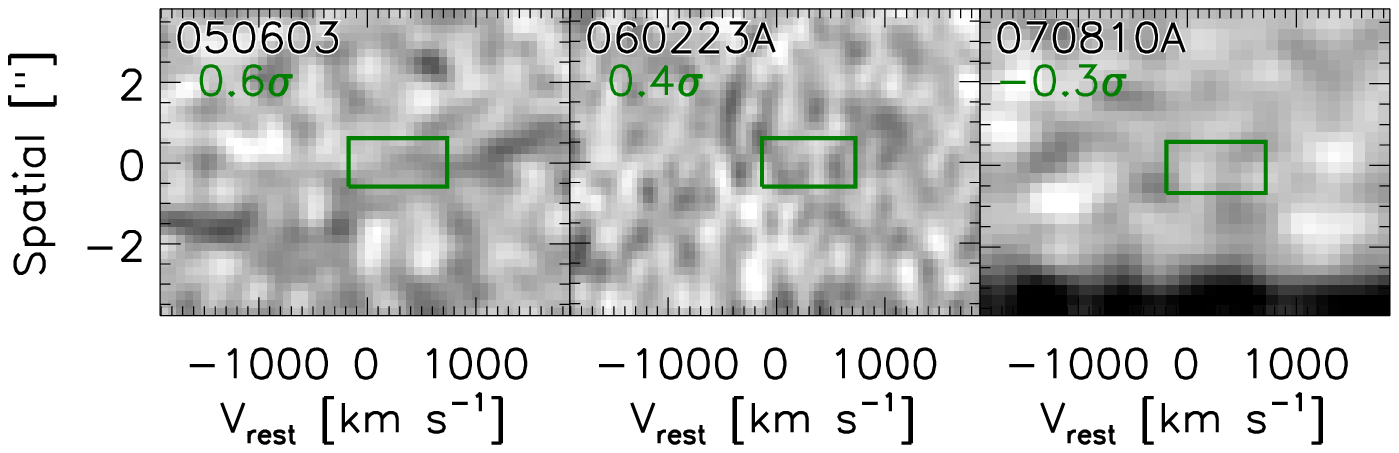}}
\caption[]{
2D spectra for the 3 systems not in the TOUGH sample.
See Fig.~\ref{fig:2Dspec_montage_sample} for further information.
\label{fig:2Dspec_montage_nonsample}
}
\end{figure*}

\begin{figure*}[htbp]
\makebox[\textwidth]{\includegraphics[width=1.00\textwidth,bb=20 625 547 751]{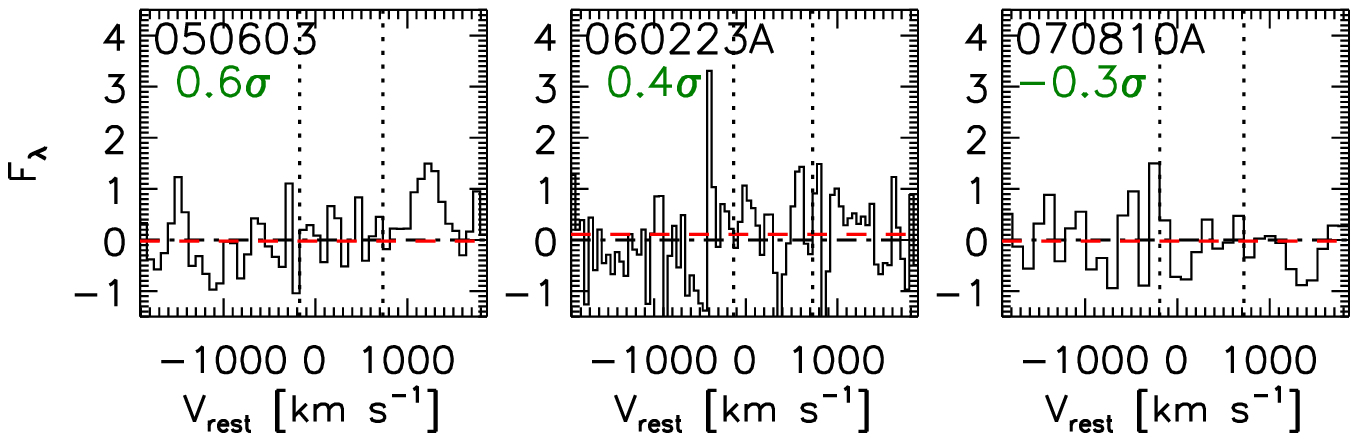}}
\caption[]{
1D spectra for the 3 systems not in the TOUGH sample.
See Fig.~\ref{fig:1Dspec_montage_sample} for further information.
\label{fig:1Dspec_montage_nonsample}
}
\end{figure*}

\begin{figure*}[htbp]
\makebox[\textwidth]{\includegraphics[width=1.00\textwidth,bb=34 625 547 751]{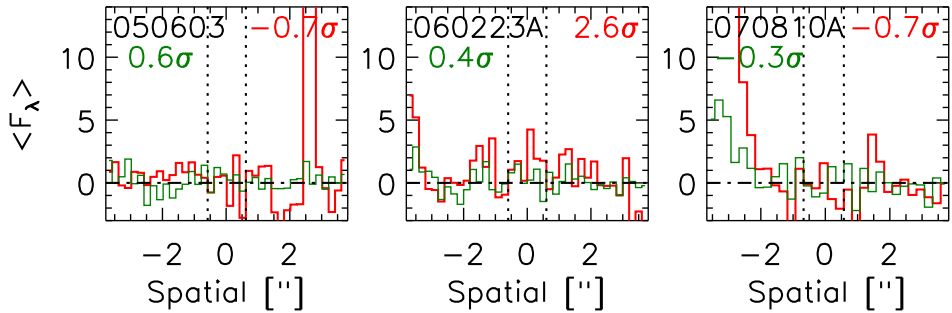}}
\caption[]{
Spatial profiles of the 3 systems not in the TOUGH sample.
See Fig.~\ref{fig:spatialprofile_montage_sample} for further information.
\label{fig:spatialprofile_montage_nonsample}
}
\end{figure*}

The spectra are shown in
Fig.~\ref{fig:2Dspec_montage_nonsample} (2D spectra),
Fig.~\ref{fig:1Dspec_montage_nonsample} (1D spectra) and
Fig.~\ref{fig:spatialprofile_montage_nonsample} (spatial profiles).
For none of these systems neither the continuum nor the {\Lya}
emission line were detected, see Table~\ref{tab:Lya_flux_nonsample}.

For GRB\,050603 the afterglow redshift of $z=2.821$ from \citet{Berger_Becker:2005}
is likely wrong: it was derived based on a reported 
bright emission line interpreted as {\Lya} in the afterglow spectrum
(0.75\,hr exposure with Magellan/IMACS),
but in our deep host spectrum (2.2\,hr exposure with VLT/FORS1)
we do not detect any emission;
we derive a $3\sigma$ upper limit on the {\Lya} flux at $z=2.821$ of
$4.7 \times 10^{-18}\,\mathrm{erg}\,\mathrm{s}^{-1}\,\mathrm{cm}^{-2}$
(cf.\ Table~\ref{tab:Lya_flux_nonsample}).
We do not find an emission line at any other redshift.

\begin{deluxetable}{lrrrrccccc}
\tabletypesize{\small}
\tablewidth{0pt}
\tablecaption{{\Lya} measurements from the spectra\label{tab:Lya_flux_nonsample}}
\tablehead{
\colhead{Name} &
\multicolumn{4}{c}{\hrulefill\ {\Lya} aperture \hrulefill} &
\colhead{$F$({\Lya})} &
\colhead{$L$({\Lya})} &
\colhead{$F_\lambda$(cont.)} &
\colhead{EW({\Lya})} &
\colhead{$v$({\Lya})} \\
\colhead{} &
\colhead{c($v$)} &
\colhead{c($s$)} &
\colhead{w($v$)} &
\colhead{w($s$)} &
\colhead{} &
\colhead{} &
\colhead{} &
\colhead{} &
\colhead{} \\
\colhead{(1)} &
\colhead{(2)} &
\colhead{(3)} &
\colhead{(4)} &
\colhead{(5)} &
\colhead{(6)} &
\colhead{(7)} &
\colhead{(8)} &
\colhead{(9)} &
\colhead{(10)}
}
\startdata
GRB\,050603  & 281 & $ 0.02$ &  906 &  1.20 & $< 9.3          $ & $<0.62          $ & $< 9.0          $                  &   \nodata                          &       \nodata   \\
GRB\,060223A & 297 & $ 0.00$ &  862 &  1.20 & $<14.5          $ & $<2.80          $ & $<12.6          $                  &   \nodata                          &       \nodata   \\
GRB\,070810A & 264 & $-0.05$ &  912 &  1.25 & $< 8.7          $ & $<0.31          $ & $< 9.5          $                  &   \nodata                          &       \nodata   \\
\enddata
\tablecomments{
See Table~\ref{tab:Lya_flux} for further information.
}
\end{deluxetable}

\end{document}